\documentclass[journal=jacsat,manuscript=article]{achemso}

\usepackage[version=3]{mhchem} 
\usepackage{caption}
\usepackage{subcaption}
\usepackage{url}
\usepackage{amsmath}
\usepackage{amssymb}
\usepackage{multirow}
\usepackage{mathrsfs}
\usepackage{booktabs}
\usepackage[english]{babel}  
\usepackage[utf8x]{inputenc}
\newcommand{\kB}{k_\mathrm{B}}

\author{Andrea Pedrielli}
\affiliation{European Centre for Theoretical Studies in Nuclear Physics and Related Areas (ECT*-Fondazione Bruno Kessler}
\alsoaffiliation{Laboratory of Bioinspired, Bionic, Nano, Meta Materials \& Mechanics, Department of Civil, Environmental and Mechanical Engineering, University of Trento, Italy}
\author{Paolo E. Trevisanutto}
\affiliation{Faculty of Engineering
Universit{\`a} Campus Bio-Medico, Rome, Italy}
\author{Lorenzo Monacelli}
\affiliation{Department of Physics, University of Rome La Sapienza, Rome, Italy}
\author{Giovanni Garberoglio}
\affiliation{European Centre for Theoretical Studies in Nuclear Physics and Related Areas (ECT*-Fondazione Bruno Kessler}
\alsoaffiliation{Trento Institute for Fundamental Physics and Applications (TIFPA-INFN), Trento, Italy}
\author{Nicola M. Pugno}
\affiliation{Laboratory of Bioinspired, Bionic, Nano, Meta Materials \& Mechanics, Department of Civil, Environmental and Mechanical Engineering, University of Trento, Italy}
\alsoaffiliation{School of Engineering and Materials Science, Queen Mary University of London, UK}
\author{Simone Taioli}
\email{taioli@ectstar.eu}
\affiliation{European Centre for Theoretical Studies in Nuclear Physics and Related Areas (ECT*-Fondazione Bruno Kessler}
\alsoaffiliation{Trento Institute for Fundamental Physics and Applications (TIFPA-INFN), Trento, Italy}
\alsoaffiliation{Peter the Great St. Petersburg Polytechnic University, St. Petersburg, Russia}

\title[An \textsf{achemso} demo]
  {Understanding Anharmonic Effects on Hydrogen Desorption Characteristics of Mg$_n$H$_{2n}$ Nanoclusters by ab initio trained Deep Neural Network}


\begin{document}







\begin{abstract}
Magnesium hydride (MgH$_2$) has been widely studied for effective hydrogen storage. However, its bulk desorption temperature ($553$~K) is deemed too high for practical applications. Besides doping, a strategy to decrease such reaction energy for releasing hydrogen is the use of MgH$_2$-based nanoparticles (NPs). Here, we investigate first the thermodynamic properties of Mg$_n$H$_{2n}$ NPs ($n<10$) from first-principles, in particular by assessing 
the anharmonic effects on the enthalpy, entropy and thermal expansion by means of the Stochastic Self Consistent Harmonic Approximation (SSCHA). The latter method goes beyond previous approaches, typically based on molecular mechanics and the quasi-harmonic approximation, allowing the ab initio calculation of the fully-anharmonic free energy. We find an almost linear dependence on temperature of the interatomic bond lengths -- with a relative variation of few percent over 300~K --, alongside with a bond distance decrease of the Mg--H bonds.
In order to increase the size of NPs toward experiments of hydrogen desorption from MgH$_2$ we devise a computationally effective Machine Learning model trained to accurately determine the forces and total energies (i.e. the potential energy surfaces), integrating the latter with the SSCHA model to fully include the anharmonic effects.
We find a significative decrease of the H-desorption temperature for sub-nanometric clusters Mg$_n$H$_{2n}$ with $n \leq 10$, with a non-negligible, although little effect due to anharmonicities (up to 10\%).
\end{abstract}

\section{Introduction}

Hydrogen represents a low environmental impact energy vector provided that effective storage media can be found \citep{Chen2005, Wu2009, Buckley2012, Shevlin2013, Hussain2015, Shen2017, Kumar2017, Zhang2017, Pasquini2018, Yartys2019, battisti2011zeolitic, garberoglio2014gas, garberoglio2012modeling, pedrielli2018gas}. Among the most promising materials, bulk magnesium hydride (MgH$_2$) presents considerable gravimetric hydrogen storage capacity ($7.6-7.7$~wt\% \citep{Schlapbach2001, SAKINTUNA2007,Webb2015, HANADA200767}), being also abundant and relatively cheap. However, the $1$~bar desorption temperature $T_d$, which is approximately $553$~K ($280^{\circ}$C) as well as the relatively slow absorption/desorption reaction kinetics hinder its possible use in real-life devices \citep{Pasquini2018}. 
Nonetheless, $T_d$ can be modified by pursuing several different routes, such as doping \citep{Kumar2017,doi:10.1063/1.4938245} and/or nanosizing of the storage materials \citep{Shevlin2013, Varin2006, AgueyZinsou2008, Paskevicius2010}.
In particular, the decreasing effect on $T_d$ of nanosizing has been recently investigated by both experiments \citep{Varin2006, AgueyZinsou2008, Paskevicius2010}, numerical simulations \citep{Wu2009, Buckley2012, Shevlin2013, Zhang2017, Wagemans2005, Kim2009, Koukaras2012,  Belyaev2016, Duanmu2016}, and analytical models \citep{Cui2018}.

In this respect, a few computational studies focused on nanoparticles (NPs) that are trimmed from the crystalline bulk, finding a structural destabilization upon decreasing the volume-to-surface ratio. 
The destabilization of MgH$_2$ NPs can be practically obtained through the synthesis of a metastable phase \citep{Shen2017,SanMartin1987,BASTIDE1980}  
-- called $\gamma$-MgH$_{2}$ -- via ball-milling. However, at odds with this finding, recent computational results indicated that the non-crystalline geometries actually increase their stability \citep{Wu2009, Shevlin2013} except for NPs containing only a few atoms. 
To carry out numerical simulations of the thermodynamic properties, in particular in MgH$_2$ NPs, a variety of theoretical and computational methods have been used, such as Diffusion Monte Carlo (DMC), Coupled Cluster CCSD(T) \citep{Wu2009, Koukaras2012}, and Density Functional Theory (DFT) \citep{Shevlin2013}. We notice that while DMC and CCSD(T) are the most accurate methods, the general trend of the $0$ K desorption energy can be also captured by DFT \citep{Shevlin2013}.
For example, DFT total energy calculations on small MgH$_2$ NPs (validated by high-accuracy CCSD(T) simulations) devised a possible explanation for their destabilization \citep{Koukaras2012}, by showing that while the desorption energy of hydrogen from crystalline samples approaches the bulk value with increasing size, the extrapolated desorption energy from amorphous NPs tends to a lower value ($54$ kJ/mol) which suits hydrogen storage applications. 
Furthermore, the $0$ K desorption energies have been assessed by means of hybrid functional DFT calculations finding a good matching with DMC and CCSD(T) calculations \citep{Shevlin2013, Koukaras2012}. 
 
We stress that the whole of these ab initio studies are focused on reckoning the properties of MgH$_2$ NPs at zero temperature, thus neglecting the effect of anharmonicity on entropy and enthalpy, and its possible impact on $T_d$. However, particularly in the case of small diameter NPs containing hydrogen, such effect can be significant \cite{Errea2014}, and thus must be accounted for.
In this respect, the anharmonic free energy is typically assessed either at classical level using force fields or from first-principles simulations within the harmonic or quasi-harmonic approximation \cite{Errea2014}. For example, DFT calculations within the harmonic approximation to assess the thermodynamic properties of bulk MgH$_2$ were reported \citep{Buckley2012}. However, within such approximation the reduction of the reaction enthalpy observed at experimental level by decreasing the NPs size cannot be reproduced \citep{Paskevicius2010}.
It is clear then that a proper assessment of the thermodynamic properties of Mg$_n$H$_{2n}$ NPs  should account for the accurate treatment of the anharmonicity.

In this work, we use first-principles simulations to determine the characteristic functions of state, such entropy and enthalpy, and the Gibbs free energy of Mg and Mg$_n$H$_{2n}$ NPs fully including the anharmonicity by using the Stochastic Self Consistent Harmonic Approximation (SSCHA). SSCHA includes both anharmonicity and zero-point energy in the calculation of the free energy at an efficient computational cost reduction with respect to third and fourth-order perturbative approaches \citep{Errea2014, Bianco2017, Monacelli2018}. We also compute $T_d$ for a number of Mg$_n$H$_{2n}$ NPs, with $n=3-10$ at different levels of accuracy. In particular, we compared the harmonic approximation with the fully anharmonic approach given by the SSCHA method. Furthermore we introduced the rotational entropy contribution in the anharmonic case, which could be relevant in the case of unsupported NPs. 

In this context, the possible enthalpy-entropy compensation has been proposed as the limiting mechanism preventing the decrease of $T_d$ \citep{Pasquini2018, Paskevicius2010}.

We also present in the Supplementary Material a thorough investigation of the frontiers orbitals (HOMO--LUMO) for the Mg$_n$H$_{2n}$ NPs computed at DFT-D3 \cite{doi:10.1063/1.3382344} level of theory.

Finally, to study realistic systems, such as those used in experiments, and derive possible trends of the thermodynamic observables with system size, we develop a Machine Learning (ML) model, which has been trained using our first-principles data to determine the forces and total energies (and then, the potential energy surfaces) of the Mg$_n$H$_{2n}$ molecular clusters, with $n \gg 10$. In fact, the free energy calculation via the SSCHA method with DFT-evaluated energies limits the number of atoms of the NPs, and one needs to devise novel approaches that make a cheaper evaluation of the free energy and of the forces without sacrificing accuracy. Indeed, by using ML we carried out the calculations of thermodynamic properties at ab initio accuracy for Mg$_n$H$_{2n}$ NPs up to $n=43$ on a laptop.
To this purpose, we have employed the SchNet–package \cite{schnet_JCP}, a continuous filter layers convolutional Neural-Network (NN) package integrating the latter in the Atomic Simulation Environment (ASE) \citep{Hjorth_Larsen_2017} 
together with the python implementation of the SSCHA code \citep{Errea2014, Bianco2017, Monacelli2018}. 

This article is organized as follows. In section 2, we describe our methodological and computational approaches to the simulation of the thermodynamic properties with the inclusion of anharmonic effects. In this section we also discuss our ML model applied to the Mg$_n$H$_{2n}$ NPs. In section 3, we use the SSCHA method for calculating the finite temperature properties of these MgH$_{2}$ NPs and of their relevant $T_d$ modified by the inclusion of anharmonicity. In section 4 we draw conclusions by summarizing our findings and discussing their applicability in further investigations.

\section{Theoretical and Computational Methods}

\subsection{Thermodynamics of hydrogen desorption}

The hydrogen desorption temperature $T_d$ in MgH$_2$ can be determined by reckoning the variation of the Gibbs free energy $\Delta G = \Delta H - T \Delta S$ as function of the pressure, where $\Delta H$ and $\Delta S$ are the changes in enthalpy and entropy of the reaction at constant temperature and volume, respectively. Indeed, at constant pressure $p=1$~bar, the relation:

\begin{equation}
    \ln{\left(\frac{p}{p_0}\right)}= - \frac{\Delta G }{RT_d}=0,
\end{equation}
\noindent
where $p_0=1$~bar and $R$ is the universal gas constant, reads:

\begin{equation}
   T_d= \frac{\Delta H(T)}{\Delta S(T)}.
\end{equation}

Since in general both $\Delta H$ and $\Delta S$ are temperature-dependent functions, $T_d$ is the value that fulfills the following implicit equation
\begin{equation}
  f(T)= T - \frac{\Delta H(T)}{\Delta S (T)} = 0.
\label{eq:Td}
\end{equation}

Enthalpy and entropy of both Mg$_n$H$_{2n}$ NPs and molecular hydrogen can be computed from ab initio simulations of $\Delta G$ at various temperatures. In particular, H$_2$ simulations were performed by considering a ortho- to para-hydrogen ratio of $3:1$. $\Delta G$ is an outcome of SSCHA calculations. While the latter method is generally used for periodic structures \cite{Errea2014, Bianco2017, Monacelli2018}, recent developments have enlarged its scope to structures with a large number of degrees of freedom and finite systems, such as NPs or large molecules.

From the knowledge of $\Delta G$, we computed the entropy $S$ at constant pressure (and of course, number of atoms) as: 
\begin{equation}\label{enthropy}
  S (T)= -\left.\frac{\partial G (T)}{\partial T}\right|_{p},
\end{equation}
while the enthalpy can be derived from:
\begin{equation}\label{enthalpy}
  H (T)= G(T) + TS(T).
\end{equation}

\subsection{Stochastic Self Consistent Harmonic Approximation}

The SSCHA method \citep{Errea2014, Bianco2017, Monacelli2018, monacelli2021} is based on a variational approach and delivers an accurate description of the phonon bands in non-perturbative regime.
The Helmholtz free energy (HFE) of a solid, which includes the contributions arising from the static lattice zero-temperature internal energy, from the thermal electronic excitation and from the ionic vibrations, reads \cite{Huang_1987}:

\begin{equation}\label{STQM}
  F_{\mathscr{H}} = \mathrm{Tr}(\rho_{\mathscr{H}} {\mathscr{H}}) + \frac{1}{\beta}\mathrm{Tr}(\rho_{\mathscr{H}} \ln{\rho_{\mathscr{H}}})
\end{equation}
\noindent where $\mathscr{H} = K  + V$ is the total Hamiltonian of the system written as the sum of kinetic energy operator ($K$) and of the many-body adiabatic potential energy ($V$) within the Born–Oppenheimer approximation. In Eq. \ref{STQM} $\rho_{\mathscr{H}} = e^{-\beta {\mathscr{H}}}/[\mathrm{Tr}(e^{-\beta {\mathscr{H}}})]$ is the density matrix, where $\beta = 1/(k_{\mathrm B} T)$, $k_{\mathrm B}$ is the Boltzmann constant, and $T$ the temperature.

The Gibbs--Bogoliubov variational principle states that for an arbitrary trial Hamiltonian  $\mathcal{H} = K + \mathcal{V}$, the HFE fulfills the following inequality:  
\begin{equation}
  F_{\mathscr{H}} \le \mathcal{F}_{\mathscr{H}}[\mathcal{H}] = F_{\mathcal{H}} + \int{d\bf{R}[\mathrm{\textit{V}} (\bf{R})-\mathcal{V}(\bf{R})]\rho_\mathcal{H}(\bf{R})}
\end{equation}
\noindent where $\bf{R}$ identifies the ion positions.
In the SSCHA method the minimization of $\mathcal{F}_{\mathscr{H}}[\mathcal{H}]$ is performed by a stochastic evaluation of the free energy and its gradient by varying the free parameters of an Hamiltonian characterised by an harmonic trial potential $\mathcal{V}$. During the minimization, the atomic coordinates are allowed to relax in order to obtain the finite temperature atomic positions.

In this work, we assume that the vibrational, rotational, and translational degrees of freedom are completely separable, and further that the MgH$_2$ nanoclusters can be considered as rigid rotors. The atomic coordinates used for the rotational entropy calculation are 
however optimized at finite temperature within the SSCHA.

Thus, the translational and rotational degrees of freedom do not enter in the simulation of the SSCHA free energy, and they are considered zero-frequency modes. However, their contribution to the free energy can be assessed from statistical mechanics owing to the relatively high temperatures under investigation.

To obtain accurate free energies (within a few meV), we progressively increased the stochastic populations up to a few thousands configurations. The calculations have been performed at $300$~K, $500$~K, and $600$~K. The free energy calculation has been extended in the range $\pm 20$~K around each calculated point, using the stochastic population generated at the reference temperature by taking advantage of the reweighting procedure \cite{Monacelli2018}. We used these data points and the associated stochastic error to derive by weighted interpolation the thermodynamic quantities as functions of temperature. 

SSCHA proceeds through a series of repeated evaluation steps, the first being the generation of the stochastic population and the calculation of the energy and forces for each element of the population, the second being the minimization of the free energy using a reweighting procedure. The resampling is performed when the stochastic population due to the reweighting procedure is no more representative of the starting one. 
Convergence studies are reported in the Supporting Information.

\subsection{Ab initio calculation of the Born-Oppenheimer potential surfaces and forces}

The SSCHA relies on the minimization of the free energy of the variational harmonic system by stochastic evaluation of the $N-$dimensional potential surface. This method starts by an ansatz on the force-constant matrix, which is typically assumed to be the harmonic one. We performed the evaluation of the potential energy, forces and dynamical matrix by means of DFT calculations, using the {\sc Quantum ESPRESSO} code suite \cite{Giannozzi2017}.

The ion-electron interaction was treated by the PBE exchange-correlation functional \cite{PhysRevLett.77.3865}. The van der Waals dispersion forces were included via the Grimme D3 semiempirical correction \citep{Grimme2006}. NPs properties were computed within a cubic cell of $2$~nm side for all structures, which is sufficient to avoid periodic image spurious interactions. Furthermore, to cutoff interactions among periodic replicas we used the Makov--Payne \citep{Makov1995} correction. With our simulation cell we obtained consistent results also using the Martyna--Tuckerman \citep{Martyna1999} correction.

To achieve fully converged energies and forces the kinetic energy cutoff of the plane wave basis set has been fixed to $80$~Ry, while the charge density cutoff four times as high. The self-consistent threshold has been set to $10^{-8}$~Ry.  The $k$-points sampling consists of the $\Gamma$-point only, owing to the finite nature of the MgH$_{2}$ nanostructures.

\subsection{Structure of Mg and MgH$_2$ nanoparticles}

To reckon $T_d$ and the equilibrium thermodynamic properties is crucial to find the atomic arrangements of the MgH$_2$ NPs in a global free energy minimum.
This task is prohibitively demanding owing to the large computational cost of determining from ab initio simulations the lowest energy configurations of clusters with an increasing number of atomic centers. 

To include the anharmonic effects in the geometry optimization of Mg$_{n}$H$_{2n}$ NPs ($n\le 10$) we start from structures corresponding to local minima of the relevant potential energy surfaces \citep{Shevlin2013, Belyaev2016, Duanmu2016}.
These geometries are then fully relaxed at finite temperature using the SSCHA method. The optimized atomic positions are selected as the centroid of the stochastic population.

To start the structural optimization of Mg$_{n}$H$_{2n}$ NPs ($n\ge 10$), we used the geometries of Ti$_n$O$_{2n}$ NPs ($n=15,20,43$) as by Ref. \citep{Hernndez2019}, on the basis of their similarity (other geometries of Ti$_n$O$_{2n}$ clusters potentially suitable can be found in Refs. \citep{LamielGarcia2017,MoralesGarca2019}).

\begin{figure}[htbp!]
\centering
\begin{subfigure}{0.09\textwidth}
\includegraphics[width=1.0\textwidth]{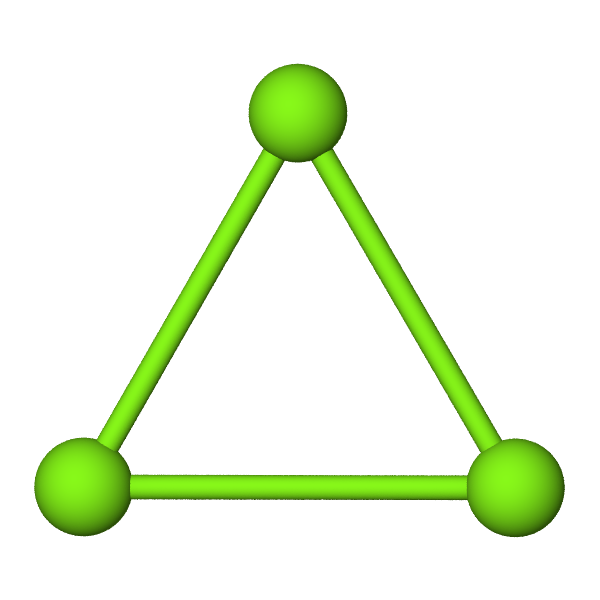}
\caption*{Mg$_3$-A}
\end{subfigure}
\begin{subfigure}{0.09\textwidth}
\includegraphics[width=1.0\textwidth]{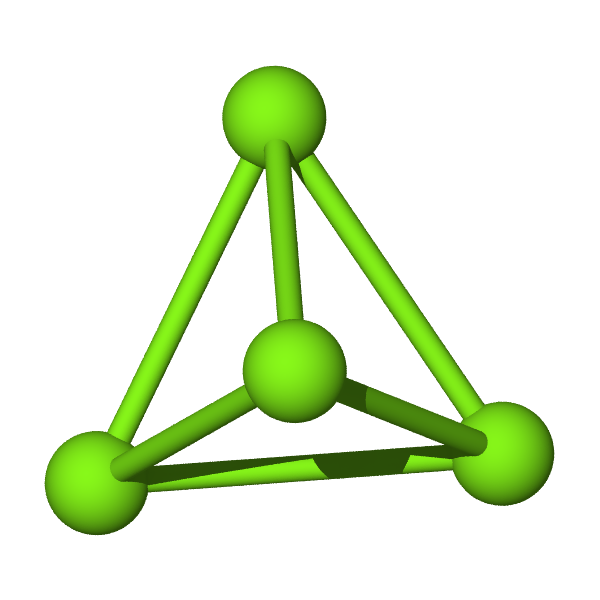}
\caption*{Mg$_4$-A}
\end{subfigure}
\begin{subfigure}{0.13\textwidth}
\includegraphics[width=1.0\textwidth]{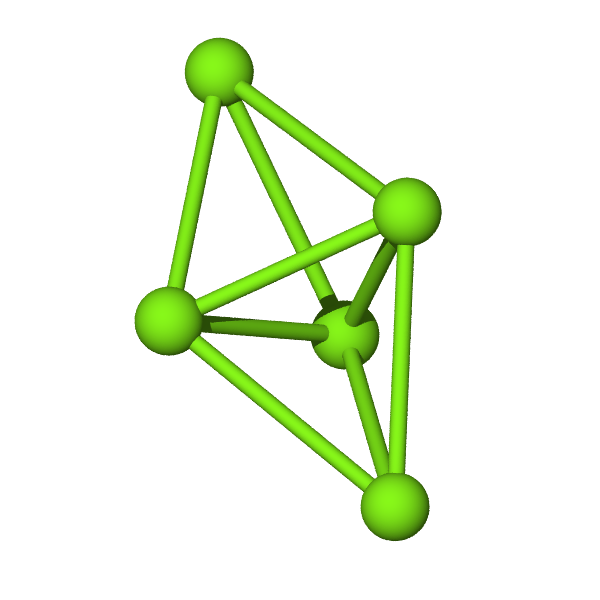}
\caption*{Mg$_5$-A}
\end{subfigure}
\begin{subfigure}{0.13\textwidth}
\includegraphics[width=1.0\textwidth]{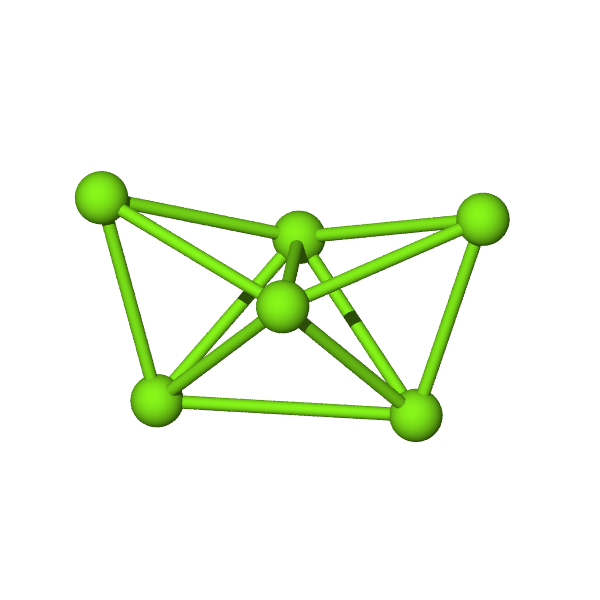}
\caption*{Mg$_6$-A}
\end{subfigure}
\begin{subfigure}{0.13\textwidth}
\includegraphics[width=1.0\textwidth]{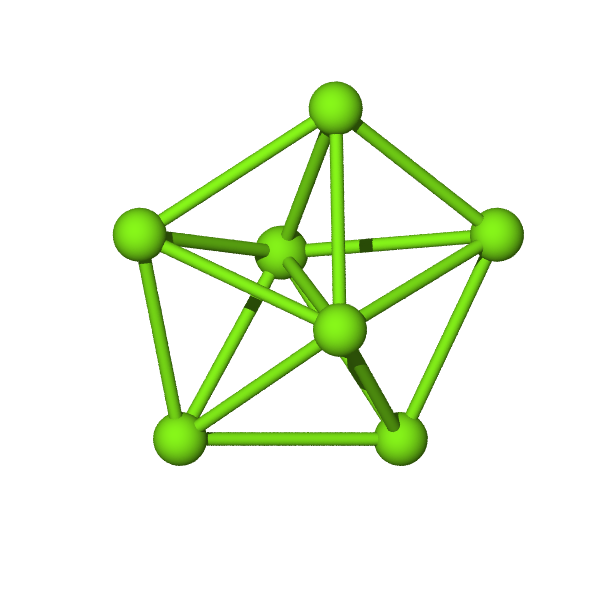}
\caption*{Mg$_7$-A}
\end{subfigure}
\begin{subfigure}{0.13\textwidth}
\includegraphics[width=1.0\textwidth]{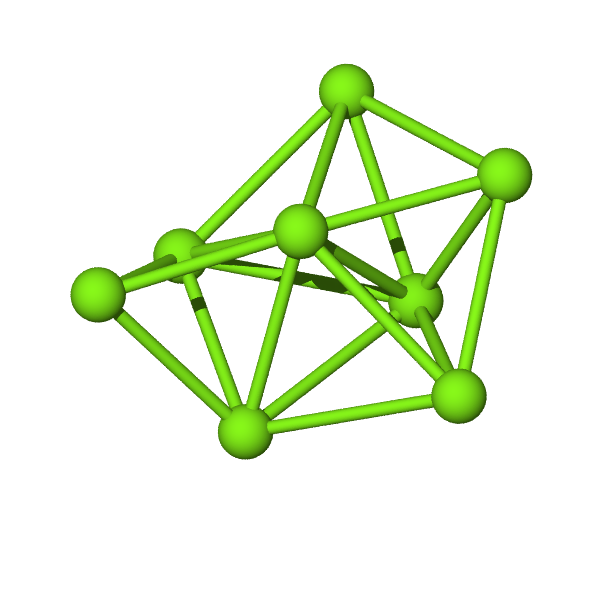}
\caption*{Mg$_8$-A}
\end{subfigure}
\begin{subfigure}{0.13\textwidth}
\includegraphics[width=1.0\textwidth]{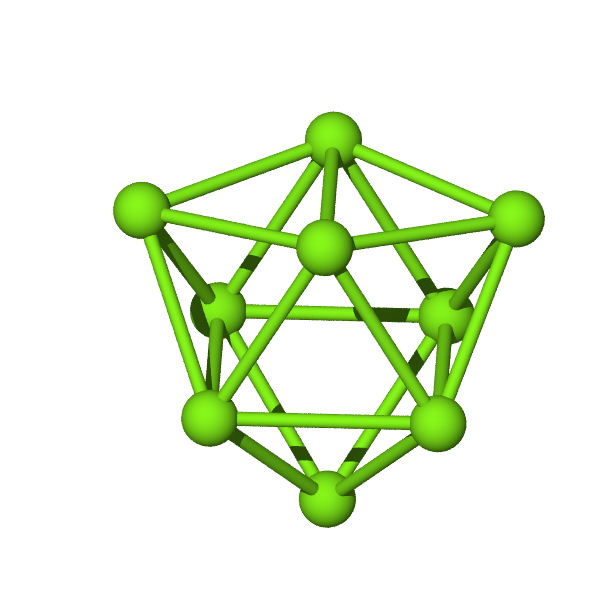}
\caption*{Mg$_9$-A}
\end{subfigure}
\begin{subfigure}{0.13\textwidth}
\includegraphics[width=1.0\textwidth]{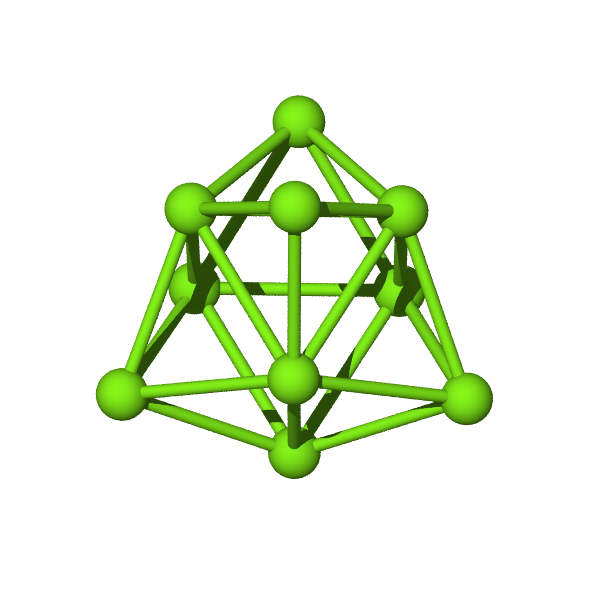}
\caption*{Mg$_{10}$-A}
\end{subfigure}
\caption{Optimized Mg$_n$ nanostructures at 300 K with the inclusion of anharmonic corrections via SSCHA.}
\label{fig:Nanoparticles1}
\end{figure}

\begin{figure}[htbp!]
\centering
\begin{subfigure}{0.13\textwidth}
\includegraphics[width=1.0\textwidth]{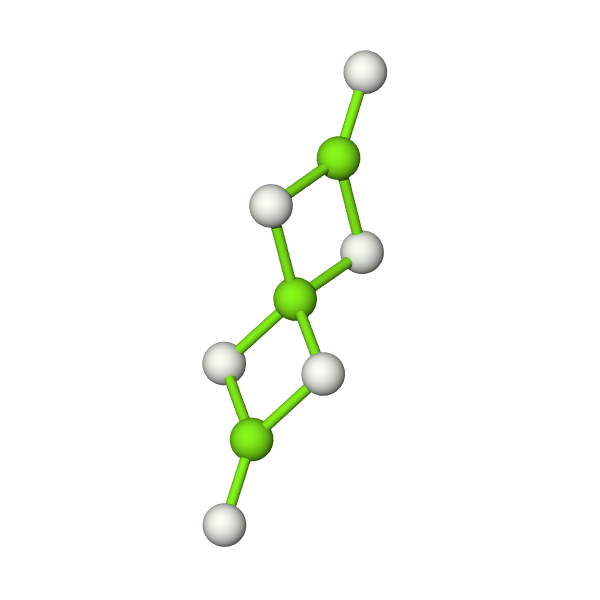}
\caption*{Mg$_3$H$_6$-A}
\end{subfigure}
\begin{subfigure}{0.13\textwidth}
\includegraphics[width=1.0\textwidth]{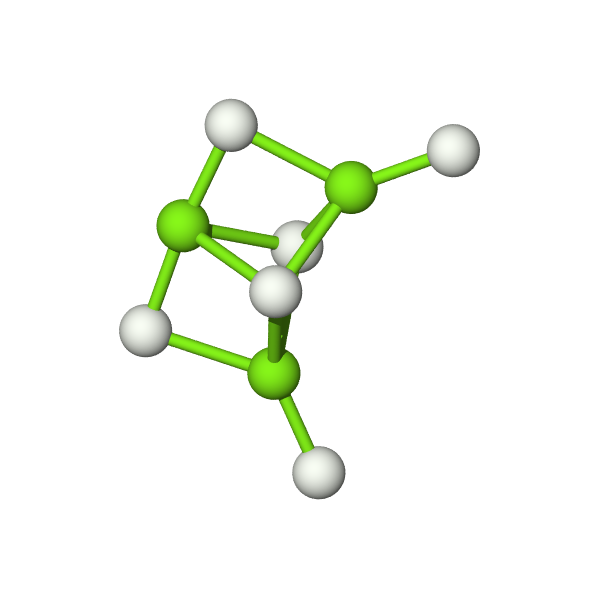}
\caption*{Mg$_3$H$_6$-B}
\end{subfigure}
\begin{subfigure}{0.13\textwidth}
\includegraphics[width=1.0\textwidth]{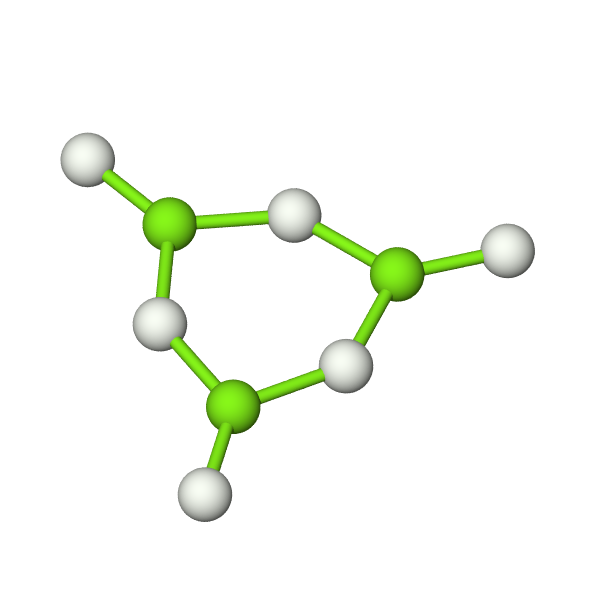}
\caption*{Mg$_3$H$_6$-C}
\end{subfigure}
\begin{subfigure}{0.13\textwidth}
\includegraphics[width=1.0\textwidth]{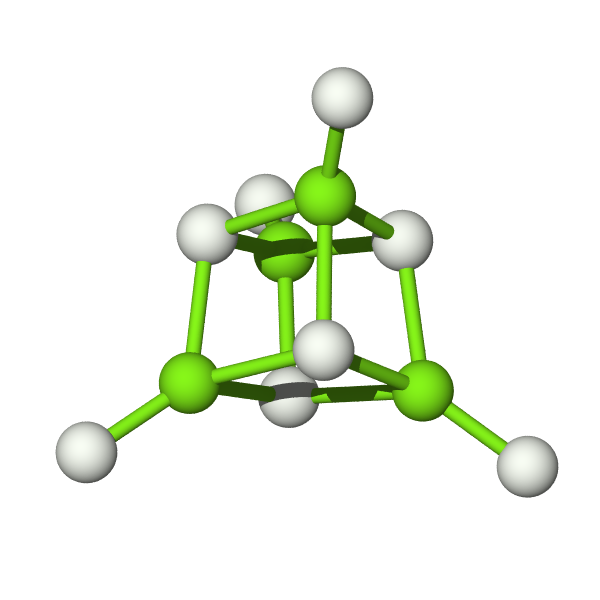}
\caption*{Mg$_4$H$_8$-A}
\end{subfigure}
\begin{subfigure}{0.14\textwidth}
\includegraphics[width=1.0\textwidth]{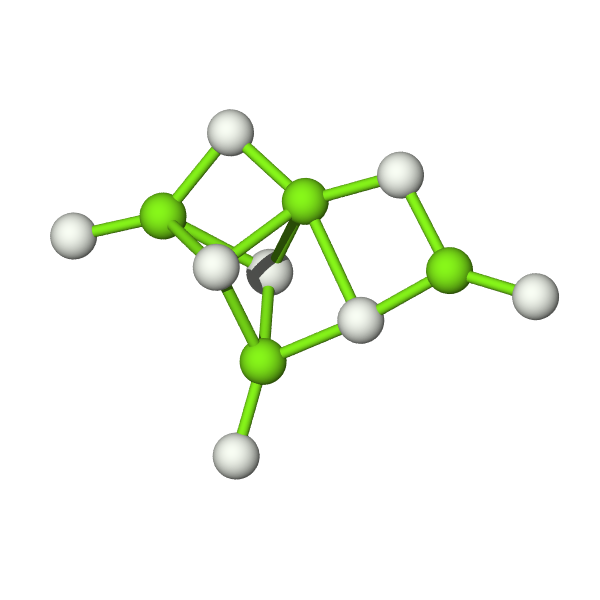}
\caption*{Mg$_4$H$_8$-B}
\end{subfigure}
\begin{subfigure}{0.14\textwidth}
\includegraphics[width=1.0\textwidth]{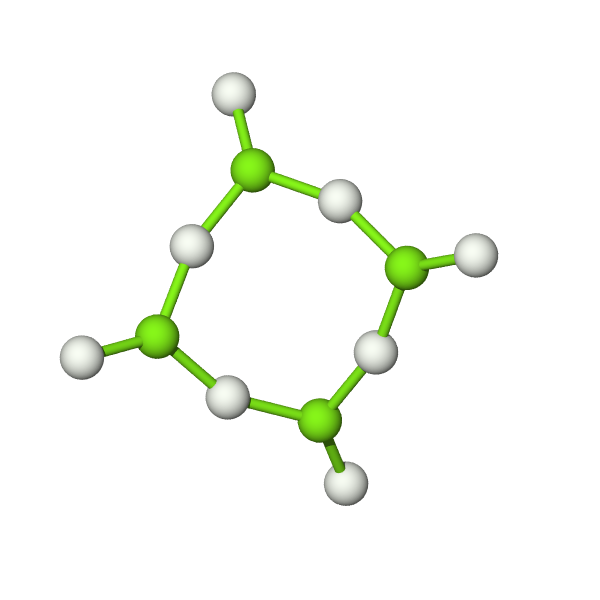}
\caption*{Mg$_4$H$_8$-C}
\end{subfigure}
\begin{subfigure}{0.14\textwidth}
\includegraphics[width=1.0\textwidth]{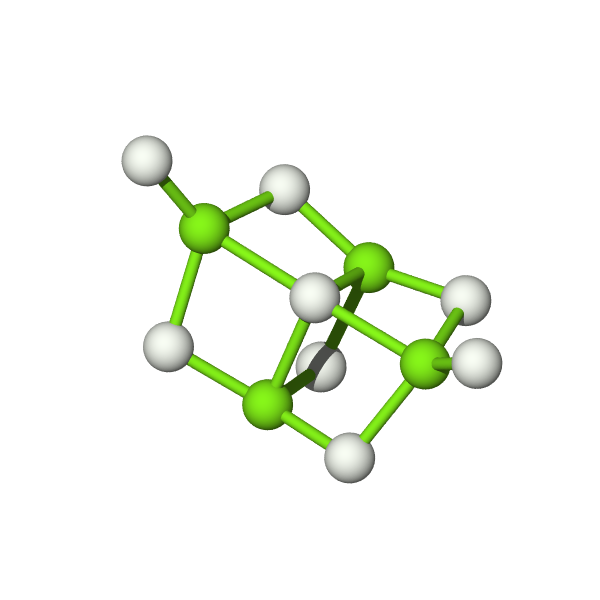}
\caption*{Mg$_4$H$_8$-D}
\end{subfigure}
\begin{subfigure}{0.14\textwidth}
\includegraphics[width=1.0\textwidth]{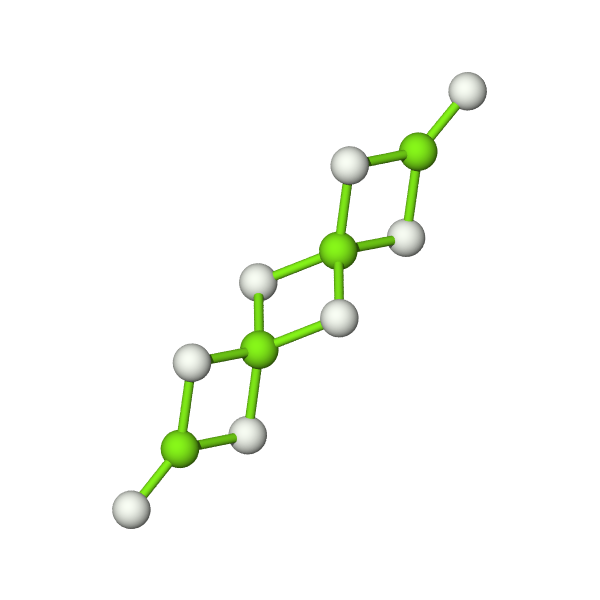}
\caption*{Mg$_4$H$_8$-E}
\end{subfigure}
\begin{subfigure}{0.14\textwidth}
\includegraphics[width=1.0\textwidth]{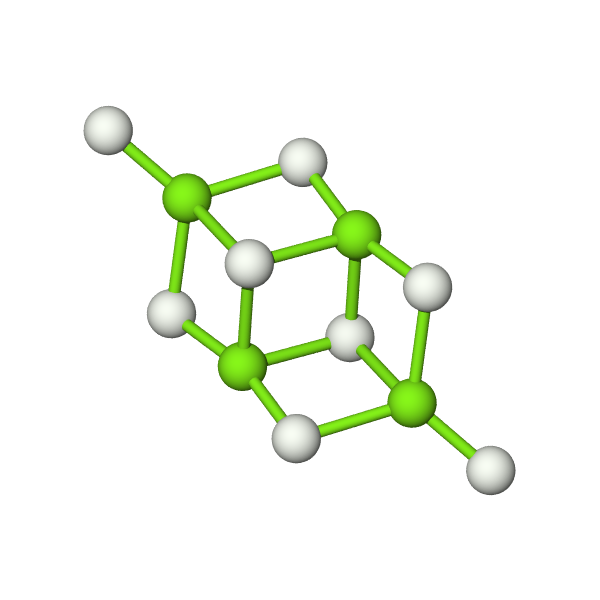}
\caption*{Mg$_4$H$_8$-F}
\end{subfigure}
\begin{subfigure}{0.14\textwidth}
\includegraphics[width=1.0\textwidth]{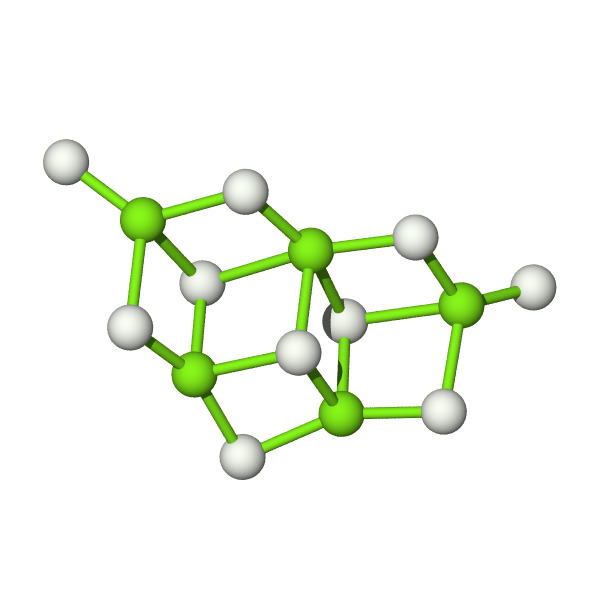}
\caption*{Mg$_5$H$_{10}$-A}
\end{subfigure}
\begin{subfigure}{0.14\textwidth}
\includegraphics[width=1.0\textwidth]{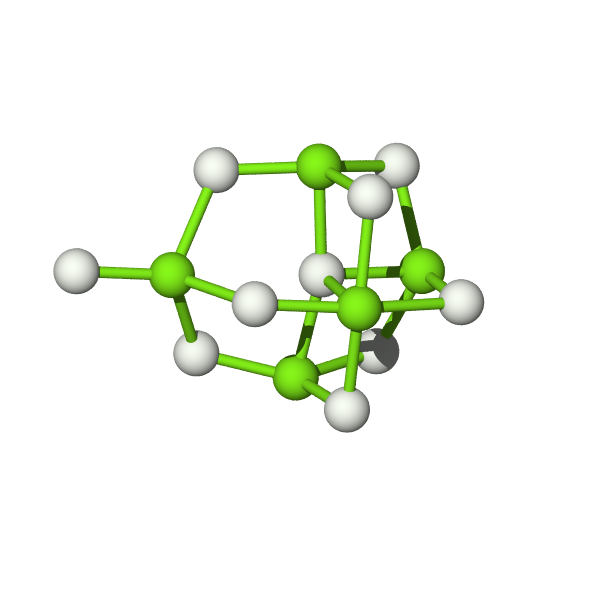}
\caption*{Mg$_5$H$_{10}$-B}
\end{subfigure}
\begin{subfigure}{0.14\textwidth}
\includegraphics[width=1.0\textwidth]{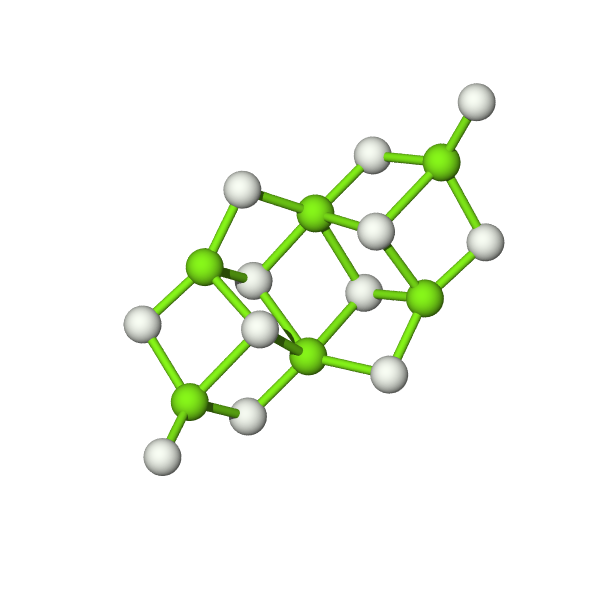}
\caption*{Mg$_6$H$_{12}$-A}
\end{subfigure}
\begin{subfigure}{0.14\textwidth}
\includegraphics[width=1.0\textwidth]{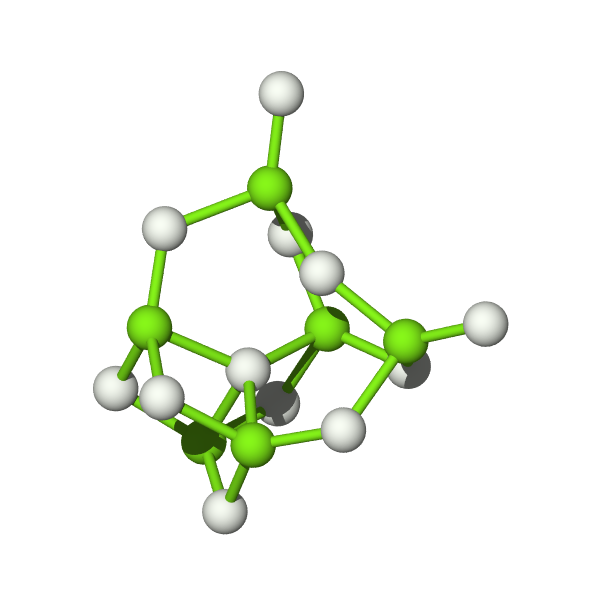}
\caption*{Mg$_6$H$_{12}$-B}
\end{subfigure}
\begin{subfigure}{0.14\textwidth}
\includegraphics[width=1.0\textwidth]{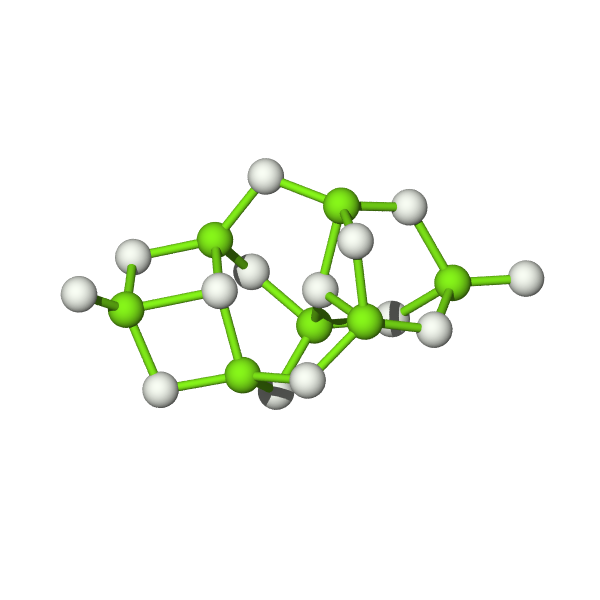}
\caption*{Mg$_7$H$_{14}$-A}
\end{subfigure}
\begin{subfigure}{0.14\textwidth}
\includegraphics[width=1.0\textwidth]{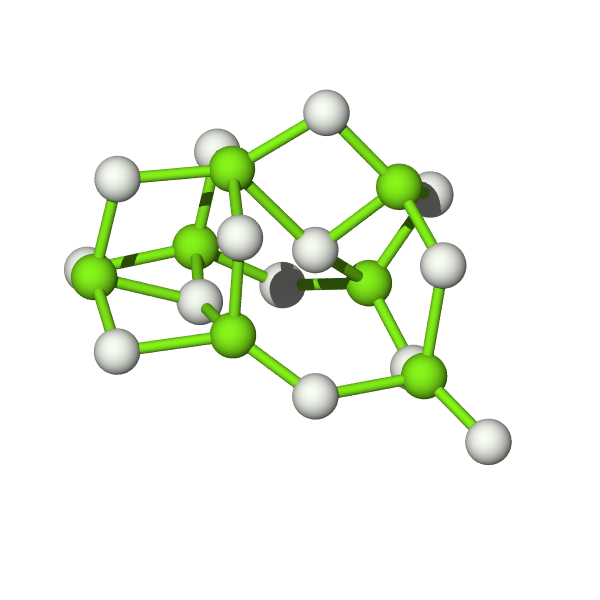}
\caption*{Mg$_7$H$_{14}$-B}
\end{subfigure}
\begin{subfigure}{0.14\textwidth}
\includegraphics[width=1.0\textwidth]{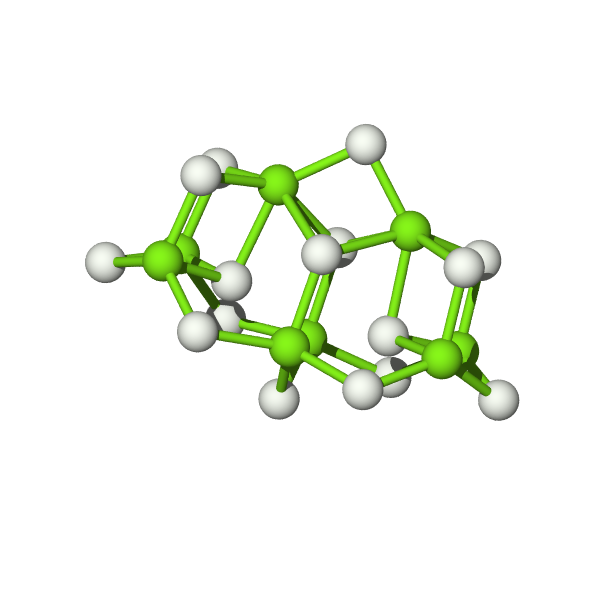}
\caption*{Mg$_8$H$_{16}$-A}
\end{subfigure}
\begin{subfigure}{0.14\textwidth}
\includegraphics[width=1.0\textwidth]{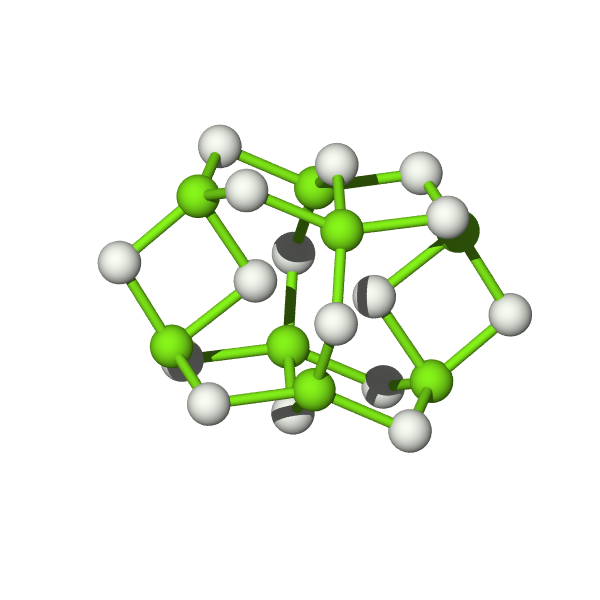}
\caption*{Mg$_8$H$_{16}$-A}
\end{subfigure}
\begin{subfigure}{0.14\textwidth}
\includegraphics[width=1.0\textwidth]{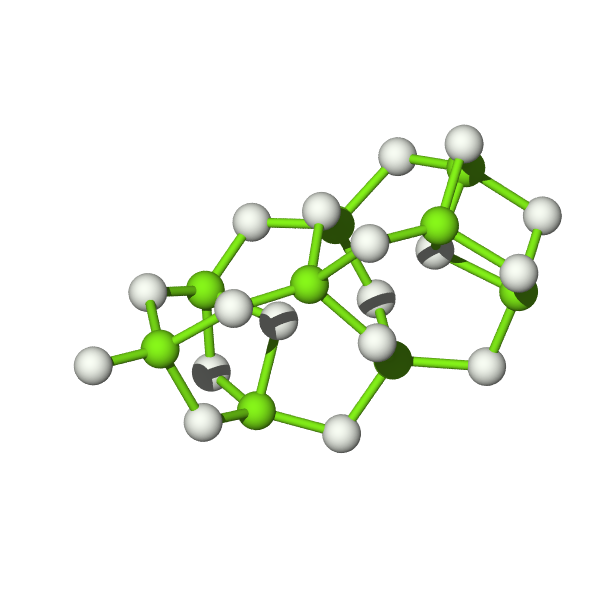}
\caption*{Mg$_9$H$_{18}$-A}
\end{subfigure}
\begin{subfigure}{0.14\textwidth}
\includegraphics[width=1.0\textwidth]{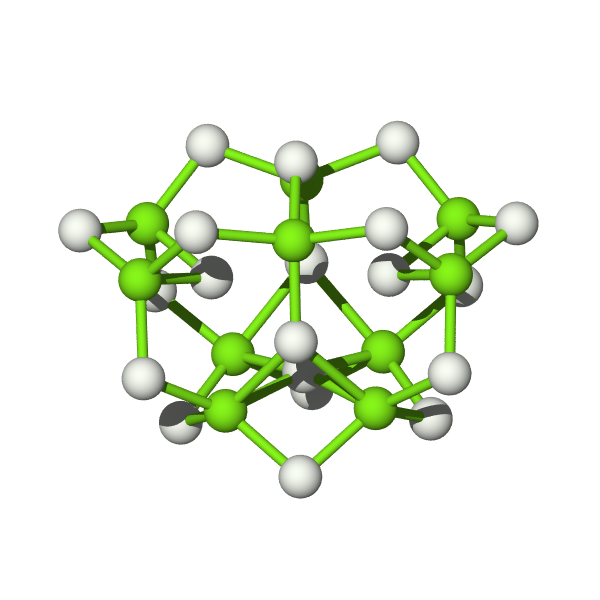}
\caption*{Mg$_{10}$H$_{20}$-A}
\end{subfigure}
\caption{Optimized MgH$_2$ nanostructures at 300 K  with the inclusion of anharmonic corrections via SSCHA.}
\label{fig:Nanoparticles2}
\end{figure}

We report in Figs. \ref{fig:Nanoparticles1} and \ref{fig:Nanoparticles2} a sketch of the $300$ K optimized Mg$_n$ and Mg$_n$H$_{2n}$ NPs ($n \le 10$), for which $T_d$ has been calculated. 

As non-stoichiometric NPs at $0$ K are less stable than stoichiometric Mg$_{n}$H$_{xn}$ clusters \citep{Shen2015} (even though substoichiometric concentration for $n>6, x=0.4$ are potentially interesting for hydrogen storage owing to their lower $T_d$ \citep{Koukaras2012}) we limit our focus to the latter ones.

\subsection{Machine Learning approach: the SchNet--SSCHA algorithm}\label{schnet}

To increase the size of MgH$_2$ NPs beyond the dimension feasible via first-principles methods, a ML model has been trained to determine the forces and the total energies (i.e. the potential energy surfaces) of the molecular clusters. The NN training set for the SchNet-SSCHA (S[chnet]SCHA) calculations is limited to energy and forces corresponding to the SSCHA populations at $300$ and $500$ K, thus is not suitable for the description of the different phases that can be crossed in the minima mining. To this purpose, we have employed SchNet \cite{schnet_JCP}, which is a continuous filter layers convolutional Neural-Network (NN) package. 

We tested our S[chnet]SCHA approach for calculating the thermodynamic properties of Mg$_n$H$_{2n}$ NPs ($n\geq 10$). In this respect, the training set contains $2 \times 10^5$ Mg$_n$ and Mg$_n$H$_{2n}$ cluster configurations ($3\le  n\le 10$) with their relevant DFT forces and total energies. The validation set contains $4 \times 10^4$ configurations. Our NN is generated with $5$ interaction blocks, 128 features and, for the Gaussian expansion, we used a range of $25$~\AA\: to cover all the interatomic distances occurring in the data. For the loss function we have also added the root-mean-square (RMS) error of the forces to the DFT total energies, with a trade-off between energy and forces loss. The convergence is obtained when the loss function is less than $10^{-4}$. A Mg$_n$ cluster test set of $2 \times 10^5$ configurations ($4 \times 10^5$ for Mg$_n$H$_{2n}$) was used for the determination of the Maximum Absolute Error (MAE) between ML predictions and the configurations of the test set. 

\begin{figure}
\includegraphics[width=0.49\linewidth]{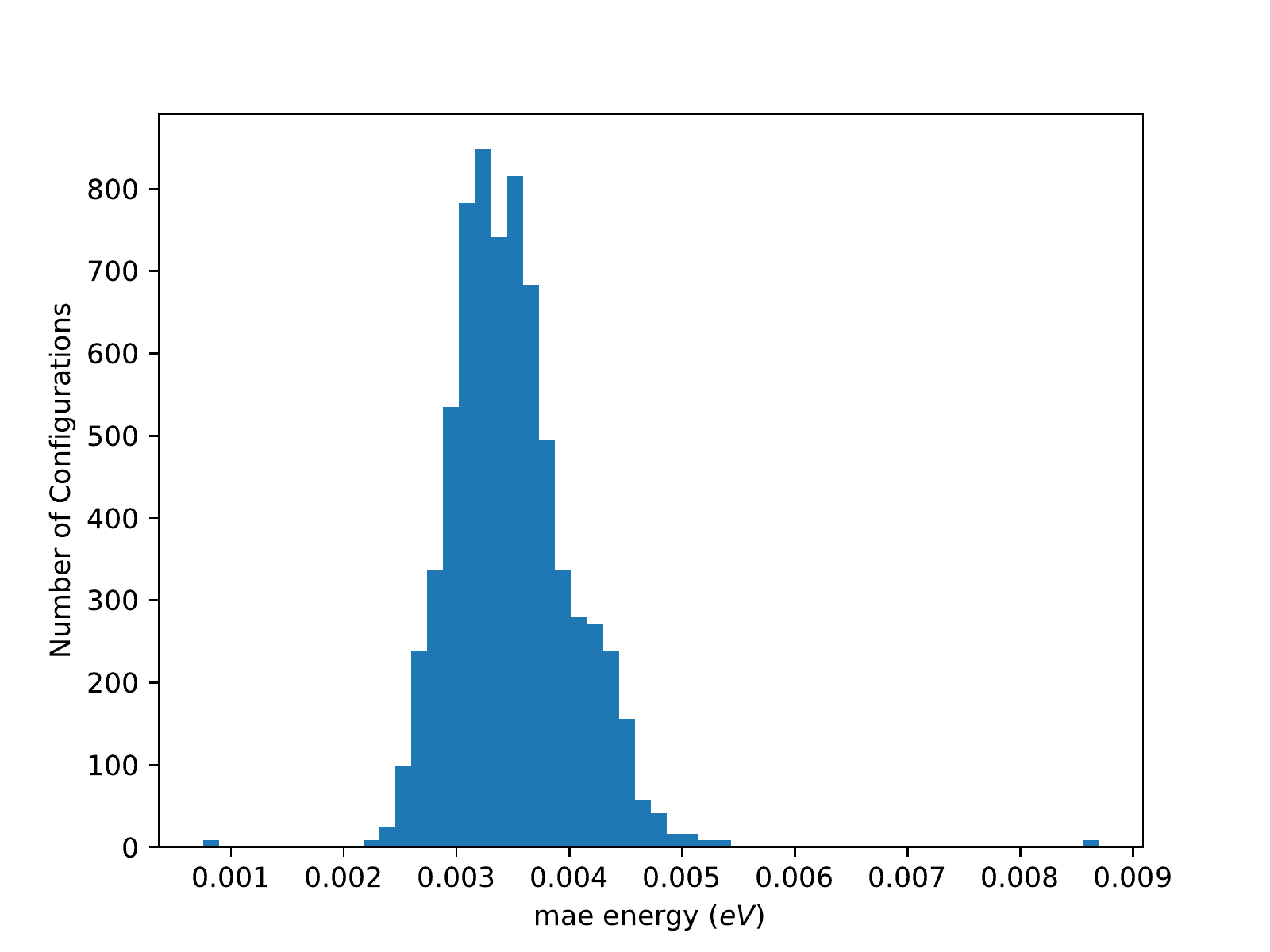} \includegraphics[width=0.49\linewidth]{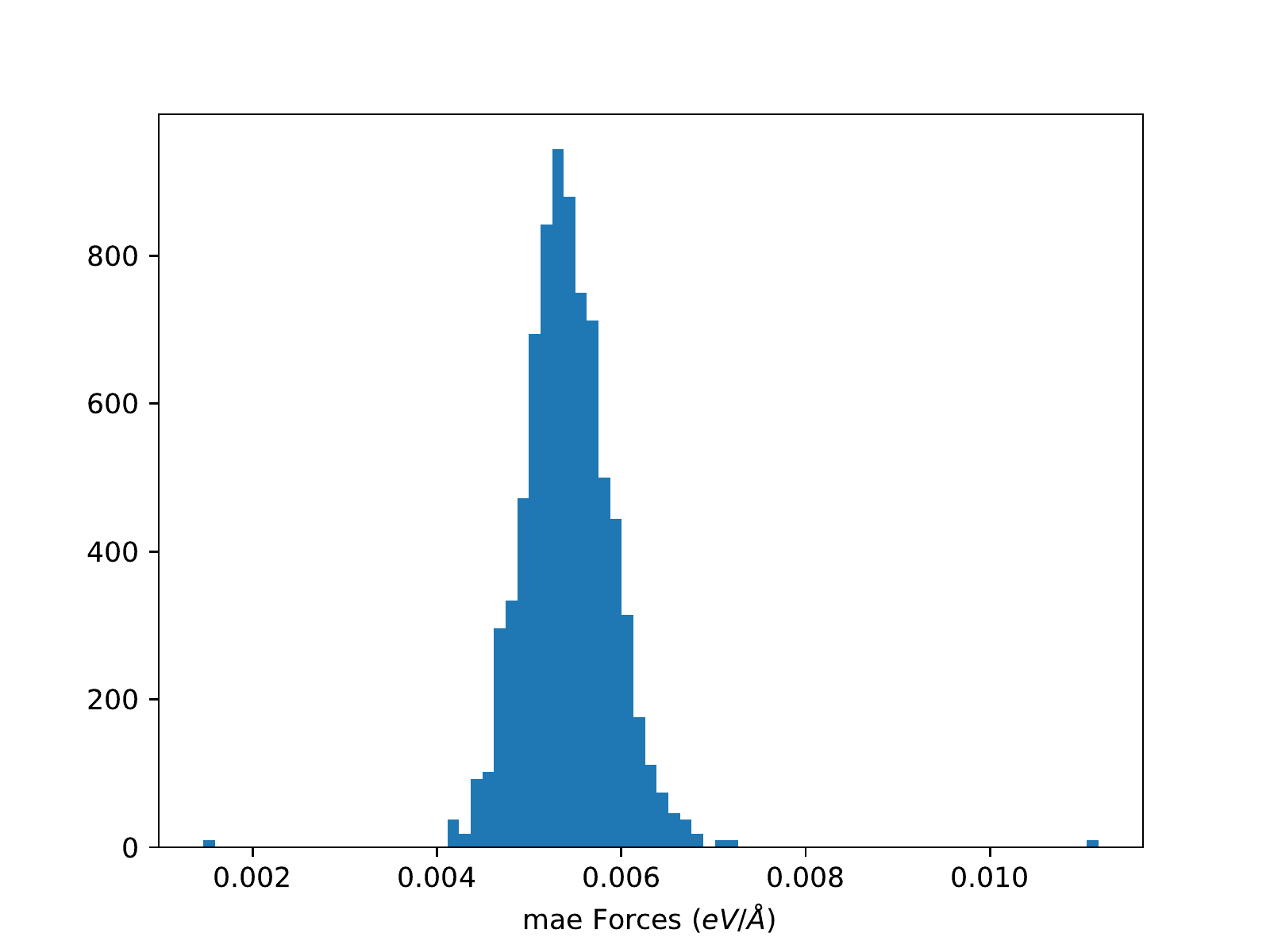}
\caption{\label{Hist_Mg}Left panel: MAE histogram distribution for the Mg$_n$ test set of the DFT total energy and the relevant ML predictions. Right panel: MAE histogram distribution for the Mg$_n$ test set of the DFT forces and the relevant ML predictions.}
\end{figure}

\begin{figure}[h]
\includegraphics[width=0.49\linewidth]{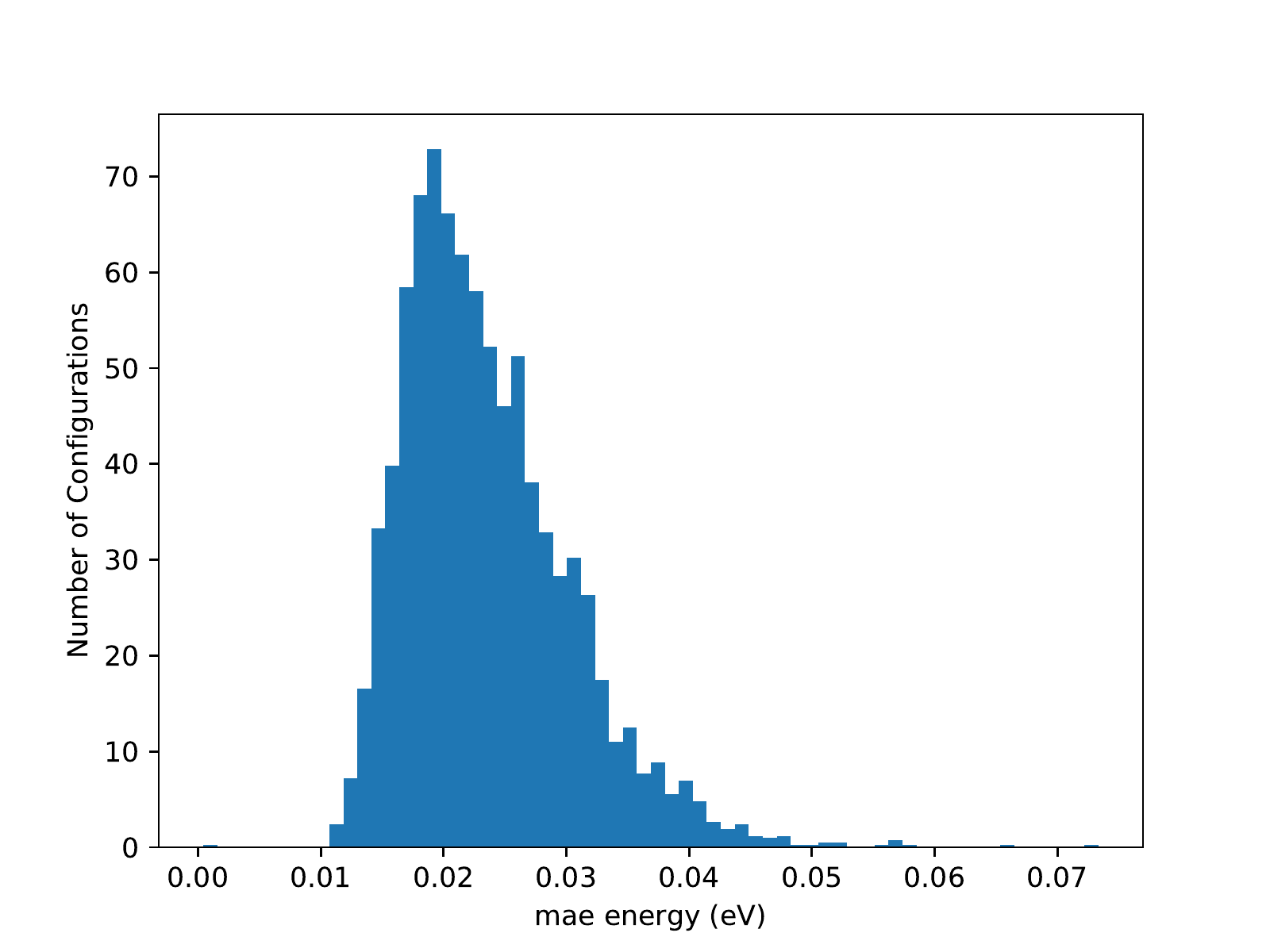} 
\includegraphics[width=0.49\linewidth,]{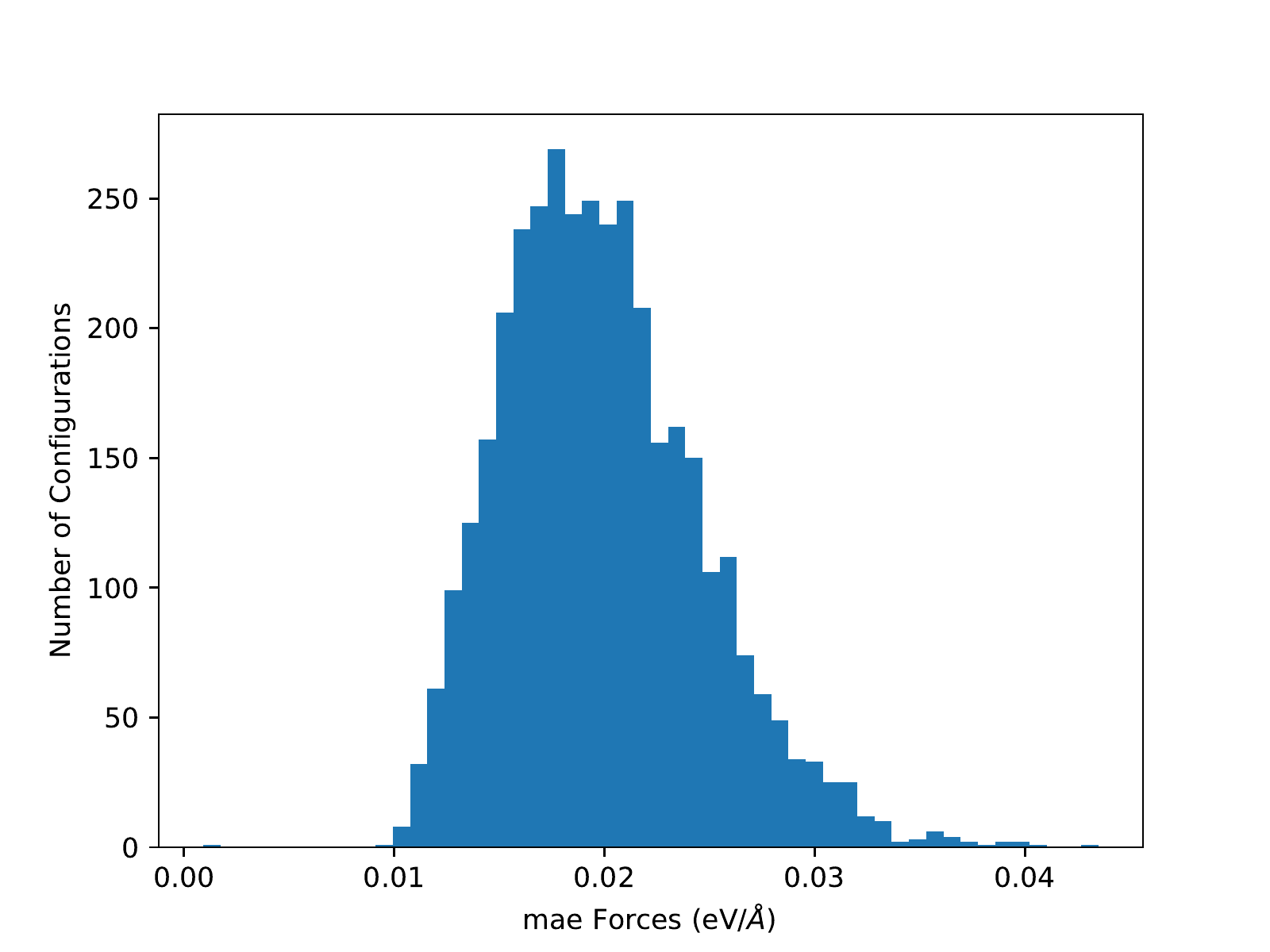}
\caption{\label{Hist_MgH2} Left panel: MAE histogram distribution for the Mg$_n$H$_{2n}$ test set of the DFT total energy and the relevant ML predictions. Right panel: MAE histogram distribution for the Mg$_n$H$_{2n}$ test set of the DFT forces and the relevant ML predictions.}
\end{figure}
In Figs. \eqref{Hist_Mg}-\eqref{Hist_MgH2}, we report the MAE histogram distributions for the ML predictions and the test set configuration batches (the batch size is made by $128$ geometrical configurations) for the case of Mg$_n$ and Mg$_n$H$_{2n}$, respectively. 

\begin{figure*}[hbtp!]
\centering
\begin{subfigure}{0.10\textwidth}
\includegraphics[width=1.0\textwidth]{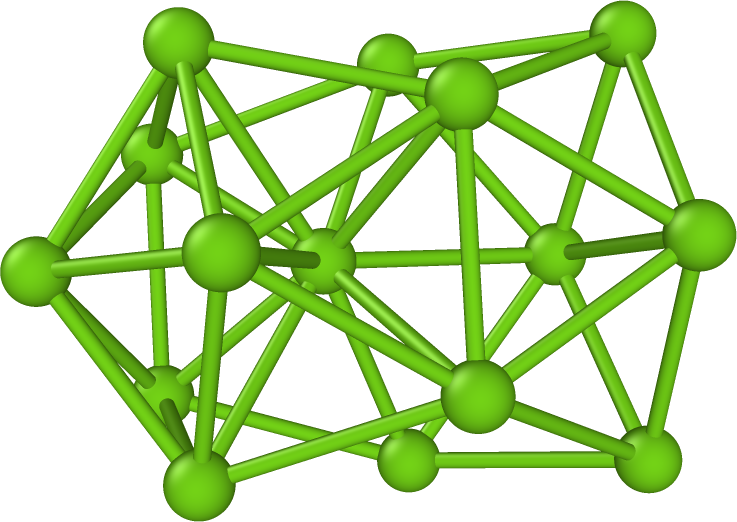}
\caption*{Mg$_{15}$}
\end{subfigure}\begin{subfigure}{0.12\textwidth}
\includegraphics[width=1.0\textwidth]{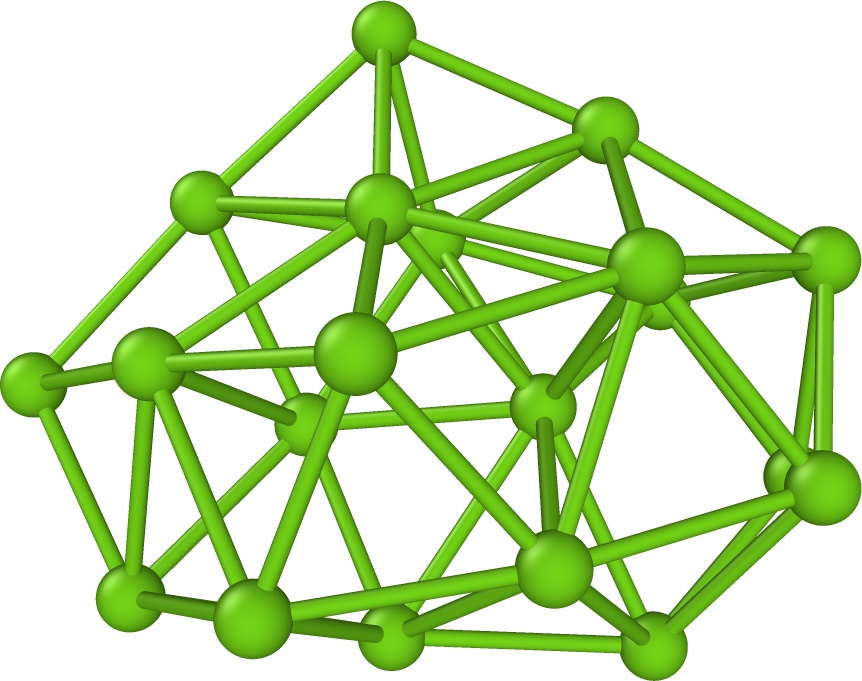}
\caption*{Mg$_{20}$}
\end{subfigure}
\begin{subfigure}{0.13\textwidth}
\includegraphics[width=1.0\textwidth]{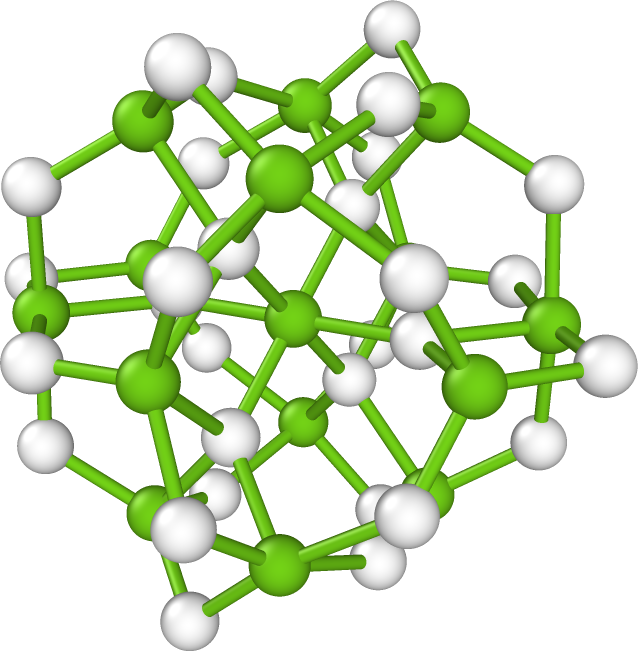}
\caption*{Mg$_{15}$H$_{30}$}
\end{subfigure}
\begin{subfigure}{0.14\textwidth}
\includegraphics[width=1.0\textwidth]{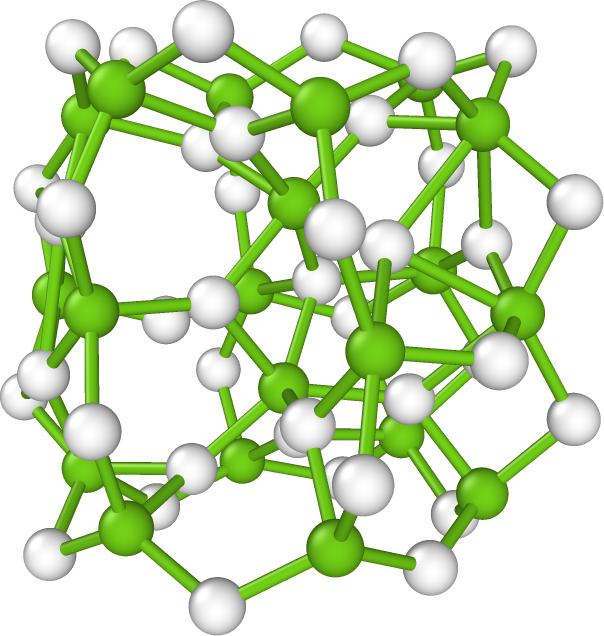}
\caption*{Mg$_{20}$H$_{40}$}
\end{subfigure}
\begin{subfigure}{0.15\textwidth}
\includegraphics[width=1.0\textwidth]{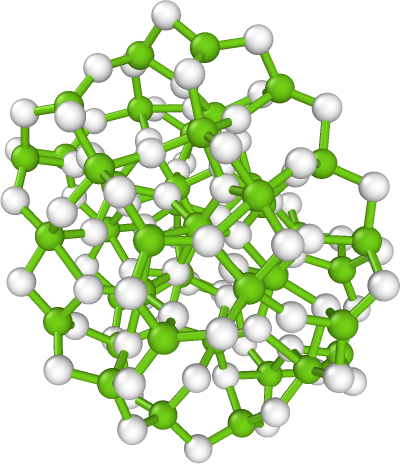}
\caption*{Mg$_{43}$H$_{86}$}
\end{subfigure}
\caption{Mg and MgH$_2$ NPs structures optimized by using our S[chnet]SCHA approach with the inclusion of anharmonic corrections.}
\label{fig:Nanoparticles3}
\end{figure*}
The MAE mean values
for the Mg$_n$ clusters are rather small, that is $0.004$ eV for the energy and $0.005$ eV/\AA~ for the forces at the peak maximum (see Fig. \eqref{Hist_Mg}). The Mg$_n$H$_{2n}$ energies and forces are, respectively, $0.02$~eV and 0.02 eV/\AA~ around the maximum of the peak distributions (see Fig. \eqref{Hist_MgH2}). 
The trained SchNet best model was embedded in the ASE calculator \cite{Hjorth_Larsen_2017}
, which provides the total energies and forces to be used in the SSCHA minimization procedure.

For $n> 10$, optimized Mg$_n$ and Mg$_n$H$_{2n}$ geometries are reported in Figs. \ref{fig:Nanoparticles3}.

\subsection{Rotational entropy calculation of Mg and MgH$_2$ nanoclusters}

Owing to the relatively high temperatures under investigation, we also assessed the contribution to the free energy of the rotational and vibrational degrees of freedom from statistical mechanics. The rotational entropy $S_{\mathrm rot}$ is expressed by \citep{Moore1963}:

\begin{equation}  
  S_{\mathrm rot}=k_{\mathrm B} \ln \left[ \frac {\sqrt{\pi I_A I_B I_C }}{\sigma} \left( \frac{8 \pi ^2 k_{\mathrm B} T }{h^2}\right)^{3/2} + \frac{3}{2}\right],
  \label{eq:RotEnt}
\end{equation}

\noindent where $I_A$, $I_B$ and  $I_C$ are the three rotational principal moments of inertia corresponding to three mutually orthogonal directions, and $h$ is the Planck constant. The symmetry number $\sigma$ takes into account the entropy reduction by dividing by the number of pure rotational symmetries of the system. In our calculations, we used $\sigma =1$ which effectively sets an upper bound for the rotational entropy of the system. 

\subsection{Calculation of the enthalpy and entropy of molecular hydrogen}

The partition function of a gas of $N$ identical and non-degenerate
molecules is given by 
\begin{eqnarray}  
  Q(V,T) = \frac{1}{N!} \left(\frac{V}{\Lambda_M^3} q(T)\right)^N ,
  \label{eq:Q}
\end{eqnarray}
where $q(T)$ is the partition function corresponding to the internal
degrees of freedom and $\Lambda_M = h / \sqrt{2 \pi M k_{\mathrm B} T}$,
is the de~Broglie thermal wavelength associated with the molecule of mass $M$.

As discussed in the supplementary information the molecular partition function can be written as
\begin{equation}
  q(T) = {\sum_L}' \sum_{n=1}^{N(L)} (2L+1) e^{-\beta E_{L,n}},
\label{eq:qfinal}  
\end{equation}
\noindent
where $E_{L,n}$ are the energies of molecular eigenstates and the primed sum denotes the sum over the angular momenta relevant to identical or distinguishable isotopes.

From $q(T)$, one can calculate, respectively \\
i) the HFE as follows:
\begin{eqnarray}
  F
  = - \kB T \ln\left( Q(V,T) \right) \nonumber \\
 = N \kB T \log\left( \frac{P \Lambda_M^3}{\kB T} \right) - N \kB T - N
  \kB T \log(q(T)), \label{eq:A}
\end{eqnarray}
ii) the average internal energy as: 
 \begin{eqnarray}
  U = -\frac{\partial \ln\left( Q(V,T) \right)}{\partial \beta} \nonumber \\
  = \frac{3}{2} N \kB T +
  N \frac{{\sum}'_{L,n} (2L+1) E_{L,N} \exp^{-\beta E_{L,n}}}{{\sum}'_{L,n}
      (2L+1) e^{-\beta E_{L,n}}},
\end{eqnarray}
iii) and, finally, the entropy and the enthalpy of the H gas as:
\begin{equation}
  S = \frac{U-F}{T}, \label{eq:S}
  \end{equation}
  \begin{equation}
 H =  U + \frac{3 N \kB T}{2} +  N \kB T, \label{eq:H}
 \end{equation}
where the ideal gas law $PV = N \kB
T$ has been used in the expression of $F$ and $H$.

The calculation of Eq.~(\ref{eq:qfinal}) has been performed by diagonalizing the Hamiltonian $\mathscr{H}$ in the subspace spanned by the states of angular momentum $L$.
The vdW corrected pair potential $V(R)$ is obtained from DFT simulations. 
We considered interatomic separations in the range $0.2 - 2.5$~\AA, discretized in $1024$ equally-spaced steps. We calculated $\mathscr{H}$ in the basis of particles confined in a box of the same extension, limiting to the first $256$ states.

The matrix diagonalization was carried out achieving an accuracy better that 1 part in $10^4$ for temperatures less than $5 \times 10^{3}$~K.
We considered progressively higher angular momenta up to $L_\mathrm{max}$, defined as the angular momentum whose ground-state energy is above
$5 \times 10^{4}$~K the minimum energy of the pair potential.
Once the values $E_{L,n}$ were obtained, the thermodynamic properties can be calculated using Eqs.~(\ref{eq:qfinal}) to (\ref{eq:S}).

\section{Results and Discussion}
\subsection{Hydrogen thermodynamics}

The calculations of entropy and enthalpy of H have been carried out according to Eqs. \ref{enthropy} and \ref{enthalpy}, respectively. We report the results in Tab. \ref{tab:H2}. The inclusion of the anharmonic correction to vibrations improves significantly the agreement with the experimental data \citep{Chase1998}, as we obtain an error $\le 0.1$\% with respect to previous estimates of $\simeq 5$\% \citep{Buckley2012}.

\begin{table}[htbp]
\centering
\begin{tabular}{lcccc}
\toprule
 & This work (\AA) & exp (\AA) \citep{Chase1998}\\
 \midrule
Bond length  &  0.746  &   0.741   \\ 
  \midrule
  & This work (meV/K) & exp (meV/K) \citep{Herzberg1959} \\
 \midrule
$S_{(298.15 K)}$ & 1.355 &  1.354   \\
  \midrule
$S_{(500 K)}$ & 1.511 & 1.510 \\
  \midrule
$S_{(600 K)}$ & 1.566 &  1.566 \\
 \midrule
  $H$  & This work (meV) & exp (meV) \citep{Chase1998} \\
 \midrule
$H_{(500 K)}$-$H_{(298.15 K)}$ &  0.0610 & 0.0609 \\
  \midrule
$H_{(600 K)}$-$H_{(500 K)}$ &  0.0304 & 0.0304 \\
\bottomrule
\end{tabular}
\caption{Bond lengths and thermodynamic properties of H$_{2}$: $S$=entropy, $H$=enthalpy in comparison to experimental data (exp) \citep{Herzberg1959,Chase1998}.}
\label{tab:H2}
\end{table}

\subsection{Thermal expansion}
\begin{figure}[htbp!]
\centering
\includegraphics[width=0.49\textwidth]{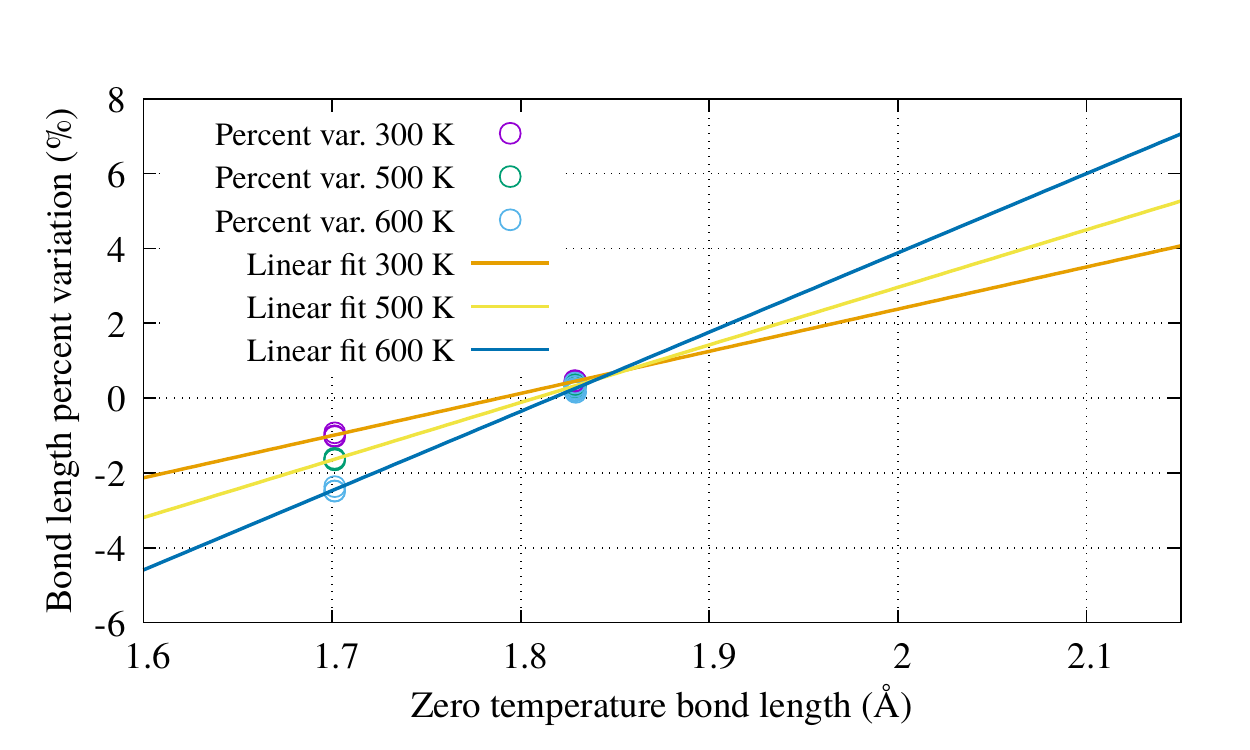}
\includegraphics[width=0.49\textwidth]{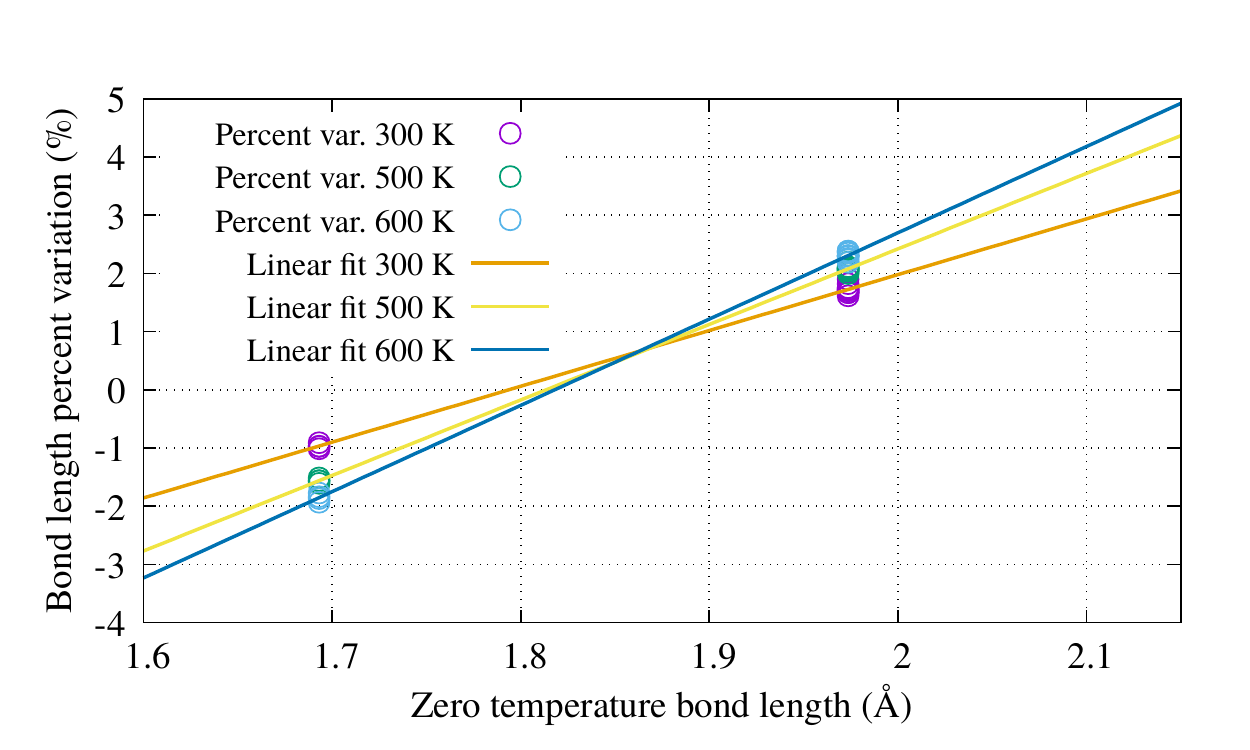}
\caption{Percent variation of 
the Mg-H bond length with respect to their 0 K values (empty circles) in the temperature range 300-600 K for the Mg$_{3}$H$_{6}$-C (left panel) and Mg$_{4}$H$_{8}$-A (right panel) NPs. The Mg-H bond length increases with temperature for those bonds where H is multi-coordinated ($\approx 1.98$~\AA ~at 0 K), and decreases for single-coordinated H atoms ($\approx 1.7$~\AA ~at 0 K). The bond lengths of double-coordinated H atoms ($\approx 1.83$~\AA ~at 0 K) are unaffected by temperature.}
\label{fig:Bonds}
\end{figure}
Starting from known local minima structures at 0 K \citep{Shevlin2013, Belyaev2016, Duanmu2016}, we optimized Mg and MgH$_2$ NPs at $300$~K using the SSCHA method, 
where the centroid positions of the stochastic population correspond to the atomic positions (see Figs. \ref{fig:Nanoparticles1} and \ref{fig:Nanoparticles2} for Mg$_n$ and Mg$_n$H$_{2n}$ NPs, respectively).
In order to explore the influence of temperature on MgH$_2$ NPs we also computed the bond lengths at $300$~K, $500$~K, and $600$~K. 

Furthermore, we reckon the percent variation of 
the Mg-H bond length with respect to their 0 K structures for all the MgH$_{2}$ NPs under investigation.
In Fig. \ref{fig:Bonds} we show the percent variation of 
the Mg-H bond length with respect to the 0 K value for Mg$_{3}$H$_{6}$-C (left panel) and of Mg$_{4}$H$_{8}$-A (right panel) NPs. These MgH$_2$ NPs clearly show that the Mg-H bond length increases with temperature for those bonds where H is multi-coordinated ($\approx 1.98$~\AA ~at 0 K) and decreases for single-coordinated H atoms ($\approx 1.7$~\AA ~at 0 K).
The bond lengths for double-coordinated H atoms ($\approx 1.83$~\AA ~at 0 K) are essentially unaffected by temperature. This picture can be extended to the other Mg$_n$H$_{2n}$ NPs, and in the Supplementary Information (see Fig. 1S) we report similar plots for various Mg$_n$H$_{2n}$ NPs in their local minima. The percent variation of the bond length due to temperature shows an approximately linear behaviour. In particular, by using a linear law to represent the percent variation $g(l)=a(l-b)$, where $l$ is the $0$~K bond length and $a,b$ are fit parameters, we find that $a$ is an increasing monotonic function of temperature, with values in the range $6$-$11$~\AA$^{-1}$, $8$-$15$~\AA$^{-1}$, and $9$-$21$~\AA$^{-1}$, for $300$~K, $500$~K, and $600$~K, respectively. The parameter $b$ is $\simeq 1.7-1.8$~\AA\:with only a small increment from $300$~K to $600$~K. 

\subsection{Thermodynamic properties} 

The thermodynamic properties per formula unit of several Mg and MgH$_2$ NPs at $300$~K obtained from our SSCHA calculations
are reported in Tabs. \ref{tab:Mg300K} and \ref{tab:MgH2300K} (simulations were also carried out at $500$~K and $600$~K).
\begin{table}[htbp!]
\centering
\begin{tabular}{lccccc}
\toprule
Nanoparticle & $U$($0$) & $F$($300$) & $F$($300$) - $U$($0$)  & $S$($300$)\\
            &  (meV)  & (meV) & (meV) & (meV/K)  \\
 \midrule
Mg$_{3}$-A	&	1516	&	1494	&	-22	&	0.153	&		\\
\midrule											
Mg$_{4}$-A	&	1331	&	1321	&	-9	&	0.175	&		\\
\midrule											
Mg$_{5}$-A	&	1314	&	1285	&	-28	&	0.258	&		\\
\midrule											
Mg$_{6}$-A	&	1291	&	1252	&	-39	&	0.308	&		\\
\midrule											
Mg$_{7}$-A	&	1224	&	1191	&	-32	&	0.314	&		\\
\midrule											
Mg$_{8}$-A	&	1187	&	1155	&	-32	&	0.319	&		\\
\midrule											
Mg$_{9}$-A	&	1096	&	1074	&	-22	&	0.288	&		\\
\midrule											
Mg$_{10}$-A	&	1027	&	1006	&	-21	&	0.290	&		\\
 \bottomrule
\end{tabular}
\caption{Thermodynamic properties of Mg NPs referred to Mg bulk formula unit at 0 K. $U$($0$) = internal energy at $0$~K, $F$($300$) and $S(300)$ are the free energy and entropy at 300~K, respectively.}
\label{tab:Mg300K}
\end{table}
 
We notice that Mg$_n$ clusters present a free energy at 300~K always lower than the internal energy at $0$~K ($F(300)-U(0)<0$), while the opposite trend was found for the Mg$_n$H$_{2n}$ NPs.
Although the result observed for Mg$_n$ does not allow to infer a significant contribution from anharmonicities (the negative sign of $F(300)-U(0)$ is generally expected due to entropy increase even in harmonic systems), the destabilization of Mg$_n$H$_{2n}$ NPs is a clear sign of the influence of anharmonicities.


Furthermore, the free energy of Mg$_n$ clusters is a decreasing function of the NPs size.
We also found that the entropy of Mg$_n$ nanoclusters shows an increasing trend up to $n=8$, while presents a slight decrease above that value. The entropy of Mg$_n$H$_{2n}$ NPs is in prevalence an increasing function of $n$. However, we notice that for a few small-$n$ Mg$_n$H$_{2n}$ NPs the entropy exceeds that of large $n$ ones.
 
\begin{table}[htbp!]
\centering
\begin{tabular}{lcccc}
\toprule
 Nanoparticle & $U$($0$) & $F$($300$) &  $F$($300$) - $U$($0$) & $S$($300$)    \\
     &   (meV)  &  (meV) & (meV) & (meV/K) \\
 \midrule
Mg$_{3}$H$_{6}$-A~(L) &	1645	&	1942	&	297	&	0.256	\\
\midrule									
Mg$_{3}$H$_{6}$-B &	1667	&	1982	&	315	&	0.191	\\
\midrule									
Mg$_{3}$H$_{6}$-C &	1733	&	2022	&	289	&	0.252	\\
\midrule									
Mg$_{4}$H$_{8}$-A &	1507	&	1816	&	310	&	0.251	\\
\midrule									
Mg$_{4}$H$_{8}$-B &	1537	&	1840	&	303	&	0.163	\\
\midrule									
Mg$_{4}$H$_{8}$-C &	1562	&	1820	&	258	&	0.355   \\
\midrule									
Mg$_{4}$H$_{8}$-D & 1416	&	1741	&	326	&	0.233	\\
\midrule									
Mg$_{4}$H$_{8}$-E &	1511	&	1806	&	296	&	0.317	\\
\midrule									
Mg$_{4}$H$_{8}$-F~(L)	 & 	1416	&	1741	&	326	&	0.237	\\
\midrule									
Mg$_{5}$H$_{10}$-A~(L) & 	1300	&	1627	&	326	&	0.269	\\
\midrule									
Mg$_{5}$H$_{10}$-B & 	1311	&	1651	&	339	&	0.236	\\
\midrule									
Mg$_{6}$H$_{12}$-A & 	1246	&	1571	&	325	&	0.297	\\
\midrule									
Mg$_{6}$H$_{12}$B~(L) & 	1184	&	1513	&	326	&	0.295   \\
\midrule									
Mg$_{7}$H$_{14}$-A~(L) & 	1094	&	1422	&	328	&	0.312	\\
\midrule									
Mg$_{7}$H$_{14}$-B  &	1119	&	1453	&	339	&	0.292	\\
\midrule									
Mg$_{8}$H$_{16}$-A~(L) &	995	&	1342	&	347	&	0.274	\\
\midrule									
Mg$_{8}$H$_{16}$-B &	1024	&	1361	&	337	&	0.295	\\
\midrule									
Mg$_{9}$H$_{18}$-A~(L) & 	967	&	1301	&	333	&	0.323	\\
\midrule									
Mg$_{10}$H$_{20}$-A	~(L) &	883	&	1224	&	341	&	0.310	\\
 \bottomrule
\end{tabular}
\caption{Thermodynamic properties of Mg$_n$H$_{2n}$ NPs referred to MgH$_2$ bulk formula unit at 0 K. $U$($0$)=internal energy at 0 K, $F$($300$) and $S(300)$ are the free energy and entropy at 300~K, respectively. (L) stands for lowest energy structure.}
\label{tab:MgH2300K}
\end{table}

\subsection{Hydrogen desorption temperature}
The first-principles calculation of the desorption temperature $T_d$, which is of paramount importance to tailor the geometrical structure and the composition of the NPs to make them viable for hydrogen storage applications, was carried out by using Eq. (\ref{eq:Td}). The dependence of $T_d$ on the cluster size was reckoned using the lowest energy Mg$_n$ and Mg$_n$H$_{2n}$ NPs for $n \le 10$.
\begin{figure}[htbp!]
\centering
\includegraphics[width=1.0\linewidth]{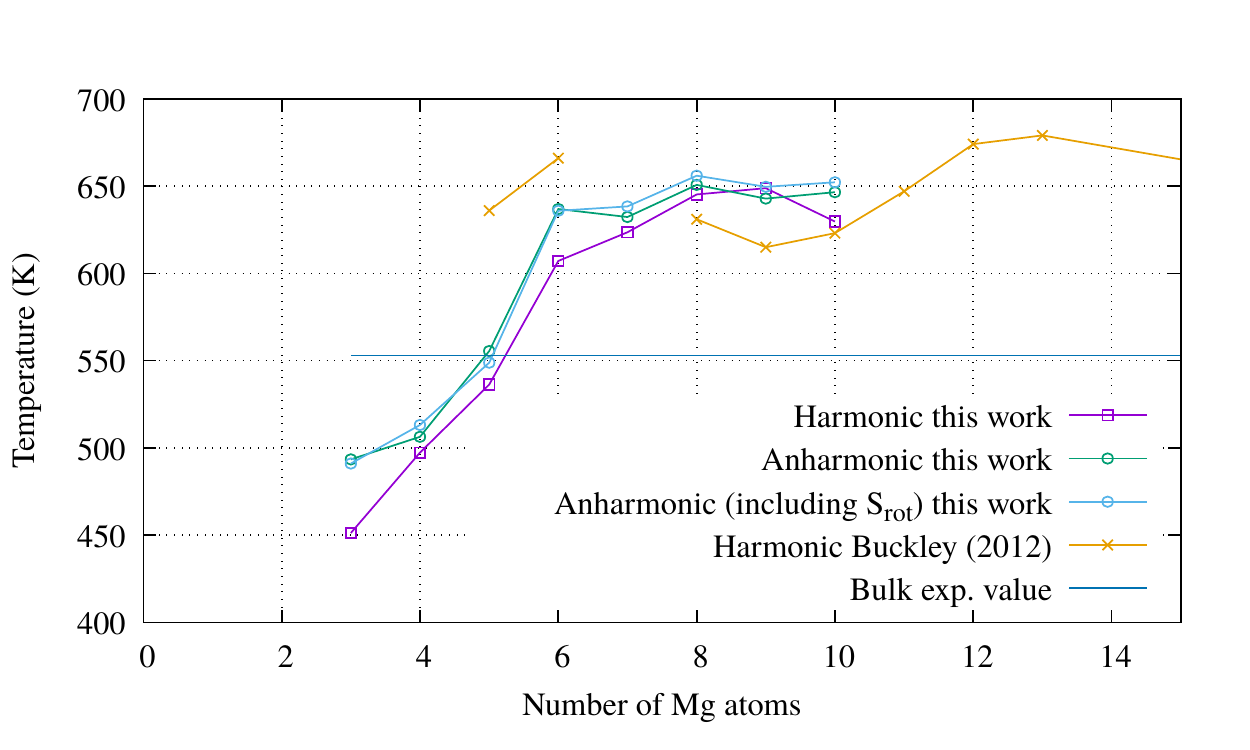}
\caption{Desorption temperature of MgH$_2$ NPs of increasing size, compared with the results obtained within the harmonic approximation \citep{Buckley2012}.}
\label{fig:Temperature}
\end{figure}

In this respect, we carried out three sets of calculations (shown in Fig.~\ref{fig:Temperature}) using i) the harmonic approximation without rotational entropy (purple line); ii) the fully anharmonic SSCHA expression neglecting the contribution of the rotational entropy (green line); iii) the fully anharmonic SSCHA expression including the rotational entropy (cyan line). In the latter case rotational and vibrational degrees of freedom were decoupled; the rotational entropy was computed including the thermal expansion of the atomic coordinates obtained from the SSCHA calculations. This reflects in the dependence of the rotational entropy (see Eq. \ref{eq:RotEnt}) on both temperature and moments of inertia. 

We report the results of these three calculations in Tab. \ref{tab:DesorptionTemperature} alongside with the desorption enthalpy and entropy for each case. Furthermore, we compare our results in Fig. \ref{fig:Temperature} with those obtained within the harmonic approximation \citep{Buckley2012} and the bulk value. 
We notice that the general trend, common to all of the three approaches, points towards a destabilization of only the smaller NPs with respect to bulk. The increase of $T_d$ with the number of atoms drives its value beyond the bulk desorption temperature. 
We also note that the difference between the harmonic approximation and the fully anharmonic calculation is small, although not negligible (up to 10\%) but presents a remarkable dependence on the number of atoms. Moreover, the inclusion of rotational entropy is essentially negligible and its contribution is almost independent of the NP size.

\begin{table*}[htbp]
\centering
\small
\begin{tabular}{ccccccccccccc}
\toprule
&& \multicolumn{3}{c}{Harmonic} && \multicolumn{3}{c}{Anharmonic} && \multicolumn{3}{c}{Anharmonic including $S_{\mathrm {rot}} $} \\
 \midrule
Reaction && T$_{d}$ & $H$(T$_{d}$) & $S$(T$_{d}$) && T$_{d}$ & $H$(T$_{d}$) & $S$(T$_{d}$) && T$_{d}$ & $H$(T$_{d}$) & $S$(T$_{d}$) \\
 Mg$_{n}$H$_{2n}$ $\xrightarrow{}$             &&   (K)   & (meV)  &  (meV/K)  &&   (K)   & (meV)  &  (meV/K) &&   (K)   & (meV)  &  (meV/K) \\
   Mg$_{n}$+(H$_{2}$)$_n$        &&  &  &    &  &  & &  &  &  \\
 \midrule
$n=3$	&&	451		&	607	&	1.35	&&	493	&	635	&	1.29 && 491 & 635 & 1.29	\\
\midrule															
$n=4$	&&	497		&	670	&	1.35   &&	506	& 684	&	1.35 && 513 & 684 & 1.33	\\	
\midrule															
$n=5$	&&	536		&	766	&	1.43	&&	555	&	775   & 	1.39 && 549 & 775 & 1.41	\\	
\midrule															
$n=6$	&&	607		&	851	&	1.40	&&  637   &   875	&   1.44 &&  636 & 876 & 1.44 \\
\midrule															
$n=7$	&&	623		&	874	&	1.40	&&	632	&	885	&	1.40 && 638 & 887 & 1.39 \\	
\midrule															
$n=8$	&&	645		&	936	&	1.45	&&	651	&	958	&	1.47 && 656 & 952 & 1.45	\\	
\midrule															
$n=9$	&&	649		&	871	&	1.34	&&	643	&	887	&	1.38 && 649 & 884 & 1.36  \\	
\midrule															
$n=10$	&&	630		&	888	&	1.41	&&	646	&	897	&	1.39 && 652  & 895 & 1.37 \\	
 \bottomrule
\end{tabular}
\caption{Harmonic and anharmonic results for desorption temperature ($T_d$), desorption enthalpy $H$(T$_{d}$) and desorption entropy $S$($T_{d}$) of MgH$_2$ NPs per unit formula at 0 K.}
\label{tab:DesorptionTemperature}
\end{table*}

Finally, our results for $T_d$ are in good agreement with those obtained in Ref. \citep{Buckley2012} for $n>8$, while for smaller nanoclusters they deviate significantly.
We observe a systematic difference at an intermediate number of atoms; we attribute it to the fact that different local minima have been used in our case. In our procedure, indeed, we start by known local minima of amorphous structures and we include the effect of finite temperatures via SSCHA, while in  Ref.~\cite{Buckley2012} the NPs are trimmed from crystalline phase and further optimized.
A main consequence of our approach is that our desorption temperature as a function of $n$ (see Fig.~\ref{fig:Temperature}) has a smoother behavior around $n=7$ than that found in Ref. \citep{Buckley2012}.


\subsection{Application of S[chnet]SCHA to larger nanoparticles}

\begin{table}[htbp!]
\centering
\begin{tabular}{l@{\hspace{2pt}}cc@{\hspace{2pt}}cc@{\hspace{2pt}}c}
\toprule
 Nanoparticle  & $U$($0$) & $F$($300$) &  $F$($300$) - $U$($0$) & $S$($300$)  & $T_d$  \\
 &(meV)  &    (meV) &  (meV) & (meV/K) & (K) \\
\midrule						
Mg$_{10}$-A	(AI) &	1027	&	1006	&	-21	&	0.290 &   \\
\midrule
Mg$_{10}$-A (ML)	&  1027	& 1006	 &	 -20    &	 0.289 &     \\
\midrule									
Mg$_{10}$H$_{20}$ (AI)	& 883 &	1226	&	343	& 0.308	&  646 	 \\
\midrule
Mg$_{10}$H$_{20}$ (ML)	&	883	&	1224	&	341	&	0.310 & 645 \\
\bottomrule
\end{tabular}
\caption{Comparison between the thermodynamic properties of the Mg$_{10}$ and Mg$_{10}$H$_{20}$ NPs computed by ab initio (AI) and our S[chnet]SCHA (ML) approaches, respectively. Thermodynamic properties refer to MgH$_{2}$ bulk formula unit at 0 K. $U$($0$) = internal energy at $0$~K, $F$($300$) and $S(300)$ are the free energy and entropy at 300~K, respectively; $T_d$= desorption temperature.}
\label{tab:MgH2_comp}
\end{table}
To compute the thermodynamic properties of NPs with $n > 10$ as typically found in experiments, we applied our S[chnet]SCHA approach to a set of Mg$_n$ ($n=15,20$) and Mg$_n$H$_{2n}$ ($n=15,20$) NPs (see the optimized configurations in Fig. \ref{fig:Nanoparticles3}). The geometries were initially optimized by using a BFGS minimization algorithm \citep{Hjorth_Larsen_2017} 
with the artificial neural network potentials. Minor changes in the atomic coordinates were found with respect to the starting configurations of the similar Ti$_n$O$_{2n}$ NPs. The dynamical matrices for these input configurations were computed using ASE \citep{Hjorth_Larsen_2017}. 
In Tab. \ref{tab:MgH2_comp} we compare the thermodynamic properties of the Mg$_{10}$ and Mg$_{10}$H$_{20}$ NPs computed by ab initio (AI) and our S[chnet]SCHA (ML) approaches, including the desorption temperature. We notice that the agreement is remarkable, thus validating our ML method.

Furthermore, in Tab. \ref{tab:MgH2_big300K} we report the thermodynamic properties at $300$ K for Mg$_n$H$_{2n}$ ($n=15,20$), that would have been very demanding with DFT-computed energies and forces. While these calculations refer to a local minimum, they only aim to show the applicability of our S[chnet]SCHA approach to carry out computational effective minimization of the free energy of these large NPs, running on a laptop.

\section{Conclusions}
In this work we analyse the effect of anharmonicity in the interatomic  potential on finite temperature properties, such as the free energy and entropy, for Mg and MgH$_{2}$ NPs using the SSCHA approach. In particular, we use ab initio simulations to find the NPs optimized geometries by relaxation of the stochastic populations centroids, corresponding to the average atomic positions. We also investigate the thermal expansion of Mg and MgH$_{2}$ NPs, finding a relative Mg-H bond length increase proportional to the $0$ K value and a decrease in the Mg-H bond length when H is single-coordinated. 
In general, the contribution of anharmonicity is small for small size NPs.
\begin{table}[htbp!]
\centering
\begin{tabular}{lccccc}
\toprule
 Nanoparticle  & $U$($0$) & $F$($300$) &  $F$($300$) - $U$($0$) & $S$($300$)    \\
 &(meV)  &    (meV) &  (meV) & (meV/K) \\
\midrule									
Mg$_{15}$	&  956	&	983 &	 27 &	0.236	\\
\midrule									
Mg$_{20}$	& 883 &	996	&  113 &	0.354	\\
\midrule
Mg$_{15}$H$_{30}$	& 797 &	1187	&	389	&	0.341	\\
\midrule									
Mg$_{20}$H$_{40}$	& 721 &   1132		&	411	&	0.383	\\
\bottomrule
\end{tabular}
\caption{Thermodynamic properties of Mg and Mg$_n$H$_{2n}$ NPs with $n>10$, per formula unit, using our S[chnet]SCHA approach. Thermodynamic properties refer to MgH$_{2}$ bulk formula unit at 0 K. $U$($0$) = internal energy at $0$~K, $F$($300$) and $S(300)$ are the free energy and entropy at 300~K, respectively.}
\label{tab:MgH2_big300K}
\end{table}

Moreover, we compute from ab initio simulations the H$_2$ $T_{d}$ in Mg$_n$H$_{2n}$ NPs ($n<10$), including such anharmonicity. We find that i) $T_{d}$ is an increasing function of the cluster size for the considered NPs; ii) the effect of anharmonicity on $T_{d}$ is relatively small for small clusters (such as Mg$_3$H$_{6}$) while is more significant ($\simeq 40$ K) for larger ones (such as Mg$_6$H$_{12}$).
The inclusion of the rotational entropy in the calculation of desorption temperature leads to almost negligible corrections. 

To extend the application of the SSCHA method to large Mg-based NPs we used the computationally expensive DFT calculations performed on several small clusters with similar bonding schemes for training a neural network to determine the forces and total energies of large nanoclusters. In particular, we adopted a Machine Learning approach based on the SchNet Neural-Network package, integrating the latter with the Atomic Simulation Environment and the Stochastic Self Consistent Harmonic Approximation python code.
We applied our S[chnet]SCHA algorithm to the calculation of thermodynamic properties of Mg$_{n}$H$_{2n}$ NPs up to $n=43$, significantly increasing the size of NPs at a very small computational cost with respect to ab initio simulations while keeping their accuracy. Indeed, the results we obtained using ab initio or ML approaches are in good agreement with respect to both thermodynamic properties and desorption temperature. This proof of concept demonstrates the viability of using ML-based methods to study the H desorption characteristics in magnesium clusters with realistic size.

Finally, our method can be adopted as high throughput screening of the thermodynamic properties and free-energy landscape exploration of large NPs, possibly relying for the training on hybrid functional DFT calculations \citep{Koukaras2012}.


\begin{acknowledgement}

The authors acknowledge Bruno Kessler Foundation (FBK) for providing unlimited access to the KORE computing facility.
P.E.T. acknowledges the Q@TN consortium for his support. This work was partially realized with the financial support of Caritro Foundation (Cassa di Risparmio di Trento e Rovereto, Italy) by the project ``{\it High-Z ceramic oxide nanosystems for mediated proton cancer therapy}''.
The research is also partially funded by the Ministry of Science and Higher Education of the Russian Federation as part of World-class Research Center program: Advanced Digital Technologies (contract No. 075-15-2020-934 dated 17.11.2020). Finally, the authors acknowledge fruitful discussions with I. Errea, M. Calandra and F. Mauri,  who also kindly provided the beta-version of the SSCHA code.

\end{acknowledgement}

\begin{suppinfo}

Further theoretical and computational procedures, convergence tests and characterization data for all new compounds. 
\end{suppinfo}


\providecommand{\latin}[1]{#1}
\makeatletter
\providecommand{\doi}
  {\begingroup\let\do\@makeother\dospecials
  \catcode`\{=1 \catcode`\}=2 \doi@aux}
\providecommand{\doi@aux}[1]{\endgroup\texttt{#1}}
\makeatother
\providecommand*\mcitethebibliography{\thebibliography}
\csname @ifundefined\endcsname{endmcitethebibliography}
  {\let\endmcitethebibliography\endthebibliography}{}

\end{document}